\pdfoutput=1 
\RequirePackage{fix-cm}
\documentclass[twocolumn]{svjour3}          
\smartqed  
\usepackage{graphicx}

\usepackage{amsmath}
\journalname{DESY Report 2014-054}

\begin{document}

\title{ Parton distribution functions at LO, NLO and NNLO with correlated uncertainties between orders}
\author{HERAFitter developers' team \and \\
P.~Belov$^{1,12}$\and
D.~Britzger$^{1}$\and
S.~Camarda$^{1}$\and 
A.M.~Cooper-Sarkar$^{2}$\and 
C.~Diaconu$^{3}$\and
J.~Feltesse$^{13}$\and
A.~Gizhko$^{1}$\and
A.~Glazov$^{1}$\and
V.~Kolesnikov$^{4}$\and
K.~Lohwasser$^{14}$\and
A.~Luszczak$^{5}$\and    
V.~Myronenko$^{1}$\and
H.~Pirumov$^{1}$\and
R.~Pla\v cakyt\. e$^{1}$\and
K.~Rabbertz$^{6}$\and    
V.~Radescu$^{1}$\and
A.~Sapronov$^{4}$\and
A.~Sch\"oning$^{10}$\and
S.~Shushkevich$^{1}$\and    
W.~Slominski$^{7}$\and    
P.~Starovoitov$^{1}$\and    
M.~Sutton$^{8}$\and    
J.~Tomaszewska$^{9}$\and    
O.~Turkot$^{1}$\and
G.~Watt$^{11}$\and 
K.~Wichmann$^{1}$ and
M.~Lisovyi$^{1}$ 
\vspace{0.5cm}
}


\institute{$^1$ DESY, Hamburg, Germany\\
$^{2}$ Department of Physics, University of Oxford, Oxford, United Kingdom \\
$^{3}$ CPPM, IN2P3-CNRS, Univ. Mediterranee, Marseille, France \\
$^{4}$ Joint Institute for Nuclear Research (JINR),Joliot-Curie 6, 141980, Dubna, Moscow Region, Russia \\
$^{5}$ T. Kosciuszko Cracow University of Technology \\
$^{6}$ Institut f\" ur Experimentelle Kernphysik, Karlsruhe, Germany \\
$^{7}$ Jagiellonian University, Institute of Physics, Ul. Reymonta 4, PL-30-059 Cracow, Poland \\
$^{8}$ University of Sussex, Department of Physics and Astronomy, Sussex House, Brighton BN1 9RH, United Kingdom \\
$^{9}$ Warsaw University of Technology, Faculty of Physics, Koszykowa 75, 00-662 Warsaw, Poland \\
$^{10}$ Physikalisches Institut, Universit\"at Heidelberg, Heidelberg, Germany \\
$^{11}$ Institute for Particle Physics Phenomenology, Durham University, Durham, DH1 3LE, United Kingdom \\
$^{12}$ Current address: Department of Physics, St. Petersburg State University, Ulyanovskaya 1, 198504 St. Petersburg, Russia\\
$^{13}$ CEA, DSM/Irfu, CE-Saclay, Gif-sur-Yvette, France \\
$^{14}$ DESY, Platanenallee 6, D–15738 Zeuthen, Germany \\
}

\date{}

\maketitle

\begin{abstract}
Sets of parton distribution functions (PDFs) of the proton are reported for the leading (LO), next-to-leading (NLO)
and next-to-next-to leading order (NNLO) QCD calculations. 
The parton distribution functions  are determined with the HERAFitter
program
using the data from the HERA experiments 
and preserving correlations between uncertainties for
the LO, NLO and NNLO PDF sets. 
The sets are used to study cross-section ratios and their uncertainties when calculated at different orders in QCD. 
A reduction of the overall theoretical uncertainty is observed if correlations between the PDF sets are taken into account
for the ratio of $WW$ di-boson to $Z$ boson production cross
sections at the LHC.
\end{abstract}

\section{Introduction}
\label{intro}
Accurate knowledge of the parton distribution functions (PDFs) of the proton 
is required for precision physics at the LHC. PDF sets are now available as
determined by several groups~\cite{Alekhin:2013nda,Ball:2012cx,mstw2008,Lai:2010vv,Aaron:2009aa,JimenezDelgado:2008hf} at leading-order (LO), next-to-lead\-ing-order
(NLO) and next-to-next-to-lead\-ing-order (NNLO) accuracy in QCD. To obtain
the cross-section predictions, the PDF
sets should be
paired with calculations of the 
coefficient functions at the  mat\-ching order of
the  accuracy. Theoretical uncertainties for the predictions
arise from both the PDF and coeffi\-cient-function uncertainties.

Most of the Standard Model processes at the LHC are calculated to NLO accuracy. The 
uncertainties due to missing higher orders for the coefficient functions are
typically determined  by varying factorisation and renormalisation scales.
This leads to large uncertainties often as large as $10\%$ of
predicted cross sections, which usually
exceed uncertainties due to the PDFs determination. For a handful of 
processes known at NNLO, 
the PDF uncertainties 
often exceed uncertainties due to missing higher orders in 
coefficient-function calculations.

The experimental precision achieved by the LHC experiments often exceeds 
the precision of theoretical calculations.  Ultimately a more complete
set of NNLO calculations should remedy the situation in future. At present,
special methods are employed to reduce theoretical uncertainties. One such method
is to measure ratios of observables which are expected to have similar
higher-order corrections. For example, the $W$ boson charge-asymmetry
measurements~\cite{Aad:2011dm,Chatrchyan:2013mza} employ almost full cancellation of
the scale uncertainties for $W^+$ compared to $W^-$ production.  However,
this cancellation is not always possible.  For example, the measurement of 
the $WW$ di-boson to $Z$ boson production cross-section ratio
performed by the CMS collaboration using $\sqrt{s}=7$~TeV data~\cite{Chatrchyan:2013yaa} 
benefits from  cancellation 
of the PDF uncertainties, but the scale uncertainties for the NLO calculation
dominate the theoretical uncertainty. While 
there is no complete NNLO calculation of the $WW$ production 
available at present, a reduction of the scale uncertainty 
for this ratio could be achieved by using NNLO calculations for the $Z$ boson
production cross section. To benefit from cancellation of the PDF uncertainties,
correlated sets at  NLO and NNLO are required in this case.

Several Monte Carlo (MC) simulation programs such as Powheg~\cite{powheg}, MC@NLO ~\cite{mcatnlo} and aMC@NLO~\cite{Frederix:2011zi} use NLO matrix-element
calculations which are matched to  parton showers. 
The parton-shower simulations are limited to 
leading-log accuracy at the moment requiring LO PDFs for consistency. 
Coherently determined, correlated LO and NLO PDF sets may be exploited for the determination
of PDF uncertainties for the experimental processes which are sensitive to the interplay of the hard-scattering matrix elements,
soft resummation and PDF content of the proton. An example of such 
process is the $W$ boson mass measurement using the charged-lepton
transverse-momentum distribution from the $W^{\pm} \to \ell^{\pm} \nu$ decay.

This paper reports a  determination of the PDFs  with correlated 
uncertainties for LO, NLO and NNLO sets. The sets are determined using the 
data from the HERA experiments~\cite{Aaron:2009aa}
and the HERAFitter analysis framework~\cite{HERAFitter,Aaron:2009kv,Aaron:2009aa}.
The experimental uncertainties
are estimated using the MC method~\cite{Jung:2009eq}
and then transformed to  eigenvector PDF sets~\cite{Pumplin:2000vx,Pumplin:2001ct}. 
The new PDF sets
are used to study correlations of the $Z$ boson production cross section
calculated at NLO and NNLO and to 
determine theoretical uncertainties for the $WW$ di-boson over $Z$ boson production 
cross-section ratio. An overall reduction of the theoretical uncertainty 
is observed.

\section{PDF analysis}
The PDF analysis reported in this paper uses the
combined HERA data~\cite{Aaron:2009aa}. These input data are accurate measurements of the inclusive 
deep-inelastic scattering (DIS) neutral- and charged-current 
cross sections combined by the H1 and ZEUS collaborations. 
The neutral-current data
cover a wide range in Bjorken $x$ and absolute 
four-momentum transfer squared, $Q^2$, 
sufficient to cover the LHC kinematics,
while the charged-current data provide information to disentangle contributions
from $u$-type and $d$-type quarks and anti-quarks at  $x>0.01$.  

This analysis is based on the open-source QCD fit 
framework as implemented in the HERAFitter program using the QCDNUM  
evolution code~\cite{Botje:2010ay} for  DGLAP evolution at 
LO, NLO and NNLO~\cite{%
Gribov:1972ri,Altarelli:1977zs,Curci:1980uw,Furmanski:1980cm,Moch:2004pa,Vogt:2004mw}. To compute DIS cross sections, 
the light-quark coefficient functions are calculated using QCDNUM in
the $\overline{MS}$ scheme~\cite{PhysRevD.8.3497} with
the renormalisation and factorisation scales set to $Q^2$. 

The heavy quarks are dynamically generated and the heavy-quark 
coefficient functions for the neutral-current $\gamma^*$ exchange  process  
 are calculated in the ge\-ne\-ral-mass
variable-flavour-number scheme (VFNS) of \cite{Thorne:1997ga,Thorne:2006qt,Thorne:2012az}
with up to five active quark flavours.
For the charged-current process, pure $Z$ exchange and $\gamma^*/Z$ interference
contributions to the neutral-current process, the heavy quarks are treated as massless. 
The NLO QCD analysis of the combined $F_2^{cc}$ data, performed by the 
H1 and ZEUS collaborations~\cite{Abramowicz:1900rp}, 
demonstrated that the preferred value of the
charm-quark-mass parameter,
 $M_c$, used in VFNS (related to the charm-quark pole mass) 
is strongly scheme dependent. This analysis
is repeated here to determine the preferred value for the NNLO heavy-quark
coefficient functions. As a cross check, an NLO analysis is repeated
first and found to reproduce the H1 and ZEUS results. The preferred
mass-parameter value at NLO (NNLO) is $M_c = 1.38$~GeV ($M_c = 1.32$~GeV) and it is used 
for the results reported in this paper. For the LO fit, the charm mass is set to $M_c = 1.38$~GeV. The bottom-quark-mass parameter is set to $4.75$ GeV
for fits at all orders.

The strong coupling constant  is set at the $Z$ boson mass $M_Z$ 
to $\alpha_S(M_Z) = 0.1184$~\cite{PDG} for both NLO and NNLO fits. 
The LO fit uses $\alpha_S(M_Z)=0.130$, 
similar to the values used in CTEQ6L~\cite{Pumplin:2002vw}, 
HERAPDF1.5LO~\cite{hera15lo}, MSTW08LO~\cite{mstw2008}  
and NNPDF2.1LO~\cite{Ball:2011uy} PDF sets.

The data included in the fit are required
to satisfy the $Q^2>Q^2_{\rm min}=7.5$~GeV$^2$ condition in order to stay in the kinematic
domain where perturbative QCD calculations can be applied. Variations
of these choices are considered as model PDF uncertainties.

The PDFs for the gluon and quark densities 
are parameterised at the input scale $Q^2_0=1.7$~GeV$^2$ as follows:
\begin{eqnarray}
  xg(x) &=& A_g x^{B_g} (1-x)^{C_g}  - A'_g x^{B'_g} (1-x)^{C'_g}\,;  \label{eq:1} \\
  x\bar{U}(x) &=& A_{\bar{U}} x^{B_{\bar{U}}}(1-x)^{C_{\bar{U}}}(1 + D_{\bar{U}}x + E_{\bar{U}}x^2)\,; \label{eq:2} \\
  x\bar{D}(x) &=& A_{\bar{D}} x^{B_{\bar{D}}}(1-x)^{C_{\bar{D}}}\,;  \label{eq:3} \\
   x u_v(x) &=& A_{u_v} x^{B_{u_v}}(1-x)^{C_{u_v}}(1 + E_{u_v}x^2)\,;  \label{eq:4} \\
   x d_v(x) &=& A_{d_v} x^{B_{d_v}}(1-x)^{C_{d_v}}(1 + D_{d_v}x)\,.  \label{eq:5}
\end{eqnarray}
Here the decomposition of the quark densities follows the one from~\cite{Aaron:2009kv} with $x\bar{U} = x\bar{u}$ and $x\bar{D} = x\bar{d} + x\bar{s}$. The contribution
of the $s$-quark density is coupled to the $d$-quark density as 
$x\bar{s} = r_s x\bar{d}$ with $r_s=1.0$, for fits at all
orders, as suggested by~\cite{Aad:2012sb},
and $x\bar{s} = x s$ is assumed.
The extra polynomial parameters $D_{d_v}, D_{\bar{U}}, E_{\bar{U}}$ are set to zero for the
central fit, however they are allowed to vary to estimate the 
parameterisation uncertainty. The normalisation of the $x u_v$ ($x d_v$)
valence-quark density, $A_{u_v}$ ($A_{d_v}$), is given by the quark-counting 
sum rule. The normalisation of the gluon density, $A_g$, is determined by
the momentum sum rule. The $x\to0$ behaviour of the $u$- and $d$-sea-quark 
density is assumed to be the same leading to two additional constraints
$B_{\bar{U}} = B_{\bar{D}}$ and  $A_{\bar{U}} =  A_{\bar{D}}/(1 + r_s)$. 
The negative term for the gluon density is suppressed at high $x$ by setting
$C'_g = 25$. 
After application of these constraints, the central fit has $13$ free parameters.  

The fit uses the $\chi^2$ definition from~\cite{Aaron:2009aa} with an additional
penalty term described in~\cite{Aaron:2012qi}.
The statistical uncertainties 
use expected instead of observed number of events. 
The data contain $114$ 
correlated systematic uncertainty sources as well as 
bin-to-bin uncorrelated systematic uncertainties. 
All systematic uncertainties are treated as multiplicative. 
The minimisation with respect to the correlated systematic uncertainty sources
is performed analytically while the minimisation with respect to PDF parameters
uses the MINUIT program~\cite{minuit}.
The central fit result is comparable to the HERAPDF1.0 set~\cite{Aaron:2009aa}. The $\chi^2$ per degree of freedom values, 
$\chi^2/N_{\rm dof}$, for the LO, NLO 
and NNLO
fits are $523/537$, $500/537$ and $498/537$, respectively. 

The PDF uncertainties arising 
from the experimental uncertainties are estimated using the MC method~\cite{Jung:2009eq}. 
The method consists in preparing a number of $N_r$ 
replicas of the data  by fluctuating the central values of the cross sections randomly within their 
statistical and systematic uncertainties taking into account  correlations. 
The uncorrelated and correlated experimental uncertainties are assumed to follow the Gaussian distribution. A set of
$1500$ replicas is prepared  and  used
as input for the LO, NLO and NNLO QCD fit.  
The fits are inspected to ensure that the minimisation has converged 
for fits at all three orders. Replicas where one of the fits has failed
are discarded. To check that this procedure does not introduce any bias, a 
study in which the non-converged fits are included has been performed.
It is found that the non-converged fits have negligible impact. A total of
$N_r = 1337$ replicas remain for which fits at all
orders have converged and they
are used for the further analysis. 

\begin{figure}
\begin{center}
\includegraphics[width=0.75\linewidth]{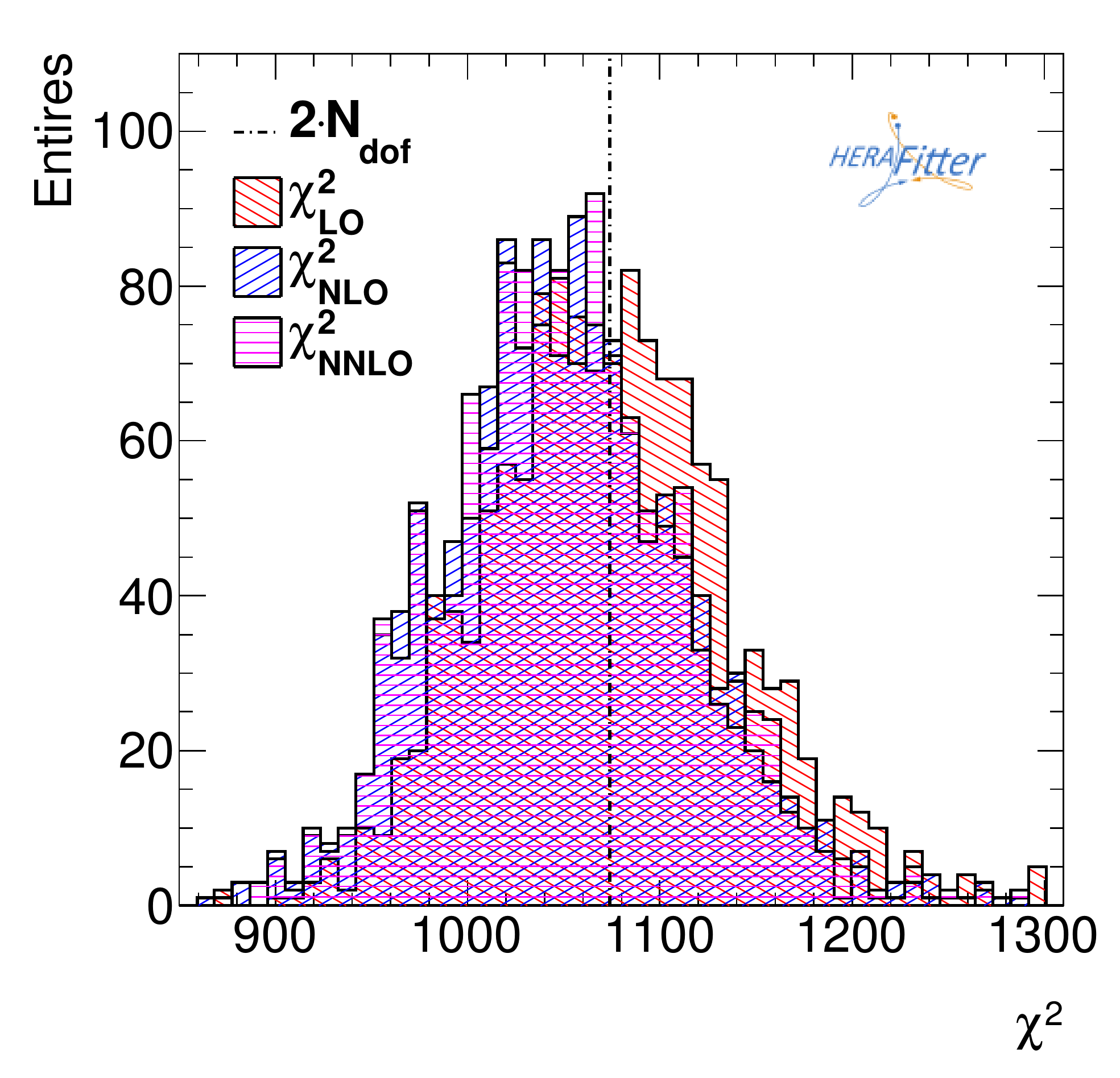}\put(-5,154){\large{(a)}}\\
\includegraphics[width=0.75\linewidth]{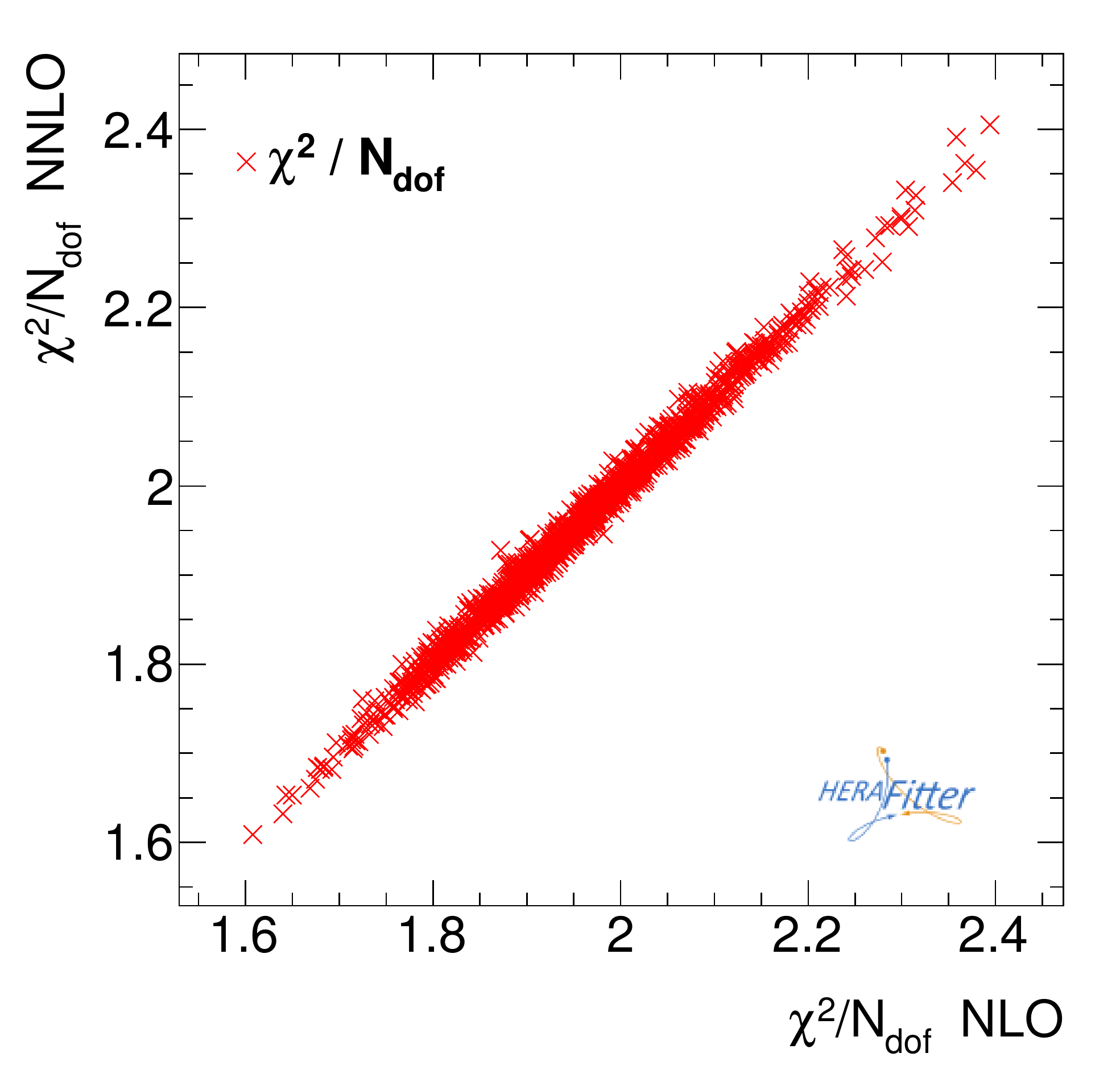}\put(-5,154){\large{(b)}}
\end{center}
\caption{\label{fig:chi2} Distribution of $\chi^2/N_{\rm dof}$ for fits to $1337$ data replicas at LO, NLO and NNLO (a). Correlation of $\chi^2/N_{\rm dof}$ 
between NLO and NNLO fits (b). The vertical line in (a) indicates the expected
mean value of the $\chi^2$ distribution for the fits to the data 
replicas in the MC method ($2\times N_{\rm dof}$). }
\end{figure}

A test of the fit results is done by investigating the $\chi^2$ distribution. For the MC method, the $\chi^2$ distribution is expected
to have a mean value of $2 N_{\rm dof}$
since it is given by the combination of  fluctuations in the data
plus random fluctuations for each MC replica. 
Figure~\ref{fig:chi2}(a) shows the observed $\chi^2$ distributions for the fits at LO, NLO and NNLO. 
The distributions follow the expected $\chi^2$ distribution.
Figure~\ref{fig:chi2}(b) shows the correlation of the $\chi^2/N_{\rm dof}$
values for the fits at NLO and NNLO. A high degree of correlation is observed.

The central values, $\mu$, 
and uncertainties, $\Delta$,
of the predictions, based on MC PDF sets, are estimated
using the mean values and standard deviations over the  predictions 
for each replica, $\sigma_i$. The predictions can be cross sections
calculated at different orders or PDFs determined at given $x,Q^2$ values.  
The correlation due to experimental uncertainties
between NLO and NNLO predictions is determined as
$$
  \begin{array}{l} \textstyle
  \rho^{\rm NLO-NNLO}   = 
    \frac{1}{N_r} \frac{ \sum^{N_r}_{i=1} (\sigma_i^{\rm NLO}-\mu^{\rm NLO} )(\sigma_i^{\rm NNLO} - \mu^{\rm NNLO}) }{ \Delta^{\rm NLO} \Delta^{\rm NNLO}}\,. 
  \end{array}
$$

\begin{figure}
\begin{center}
\hspace*{-1.4cm}\begin{minipage}{1.01\linewidth}
\begin{tabular}{cc}
\includegraphics[width=0.5\linewidth]{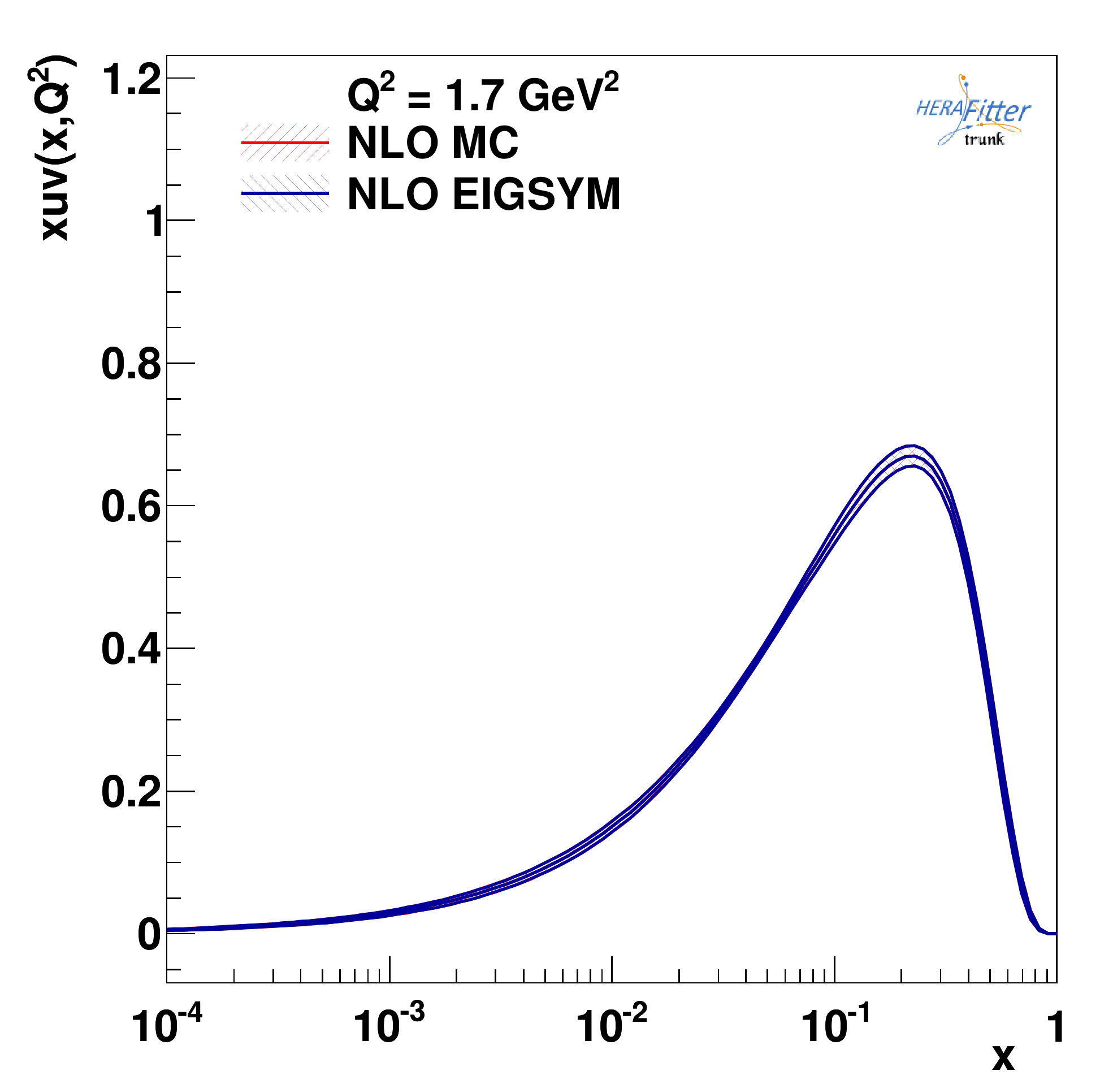} &
\includegraphics[width=0.5\linewidth]{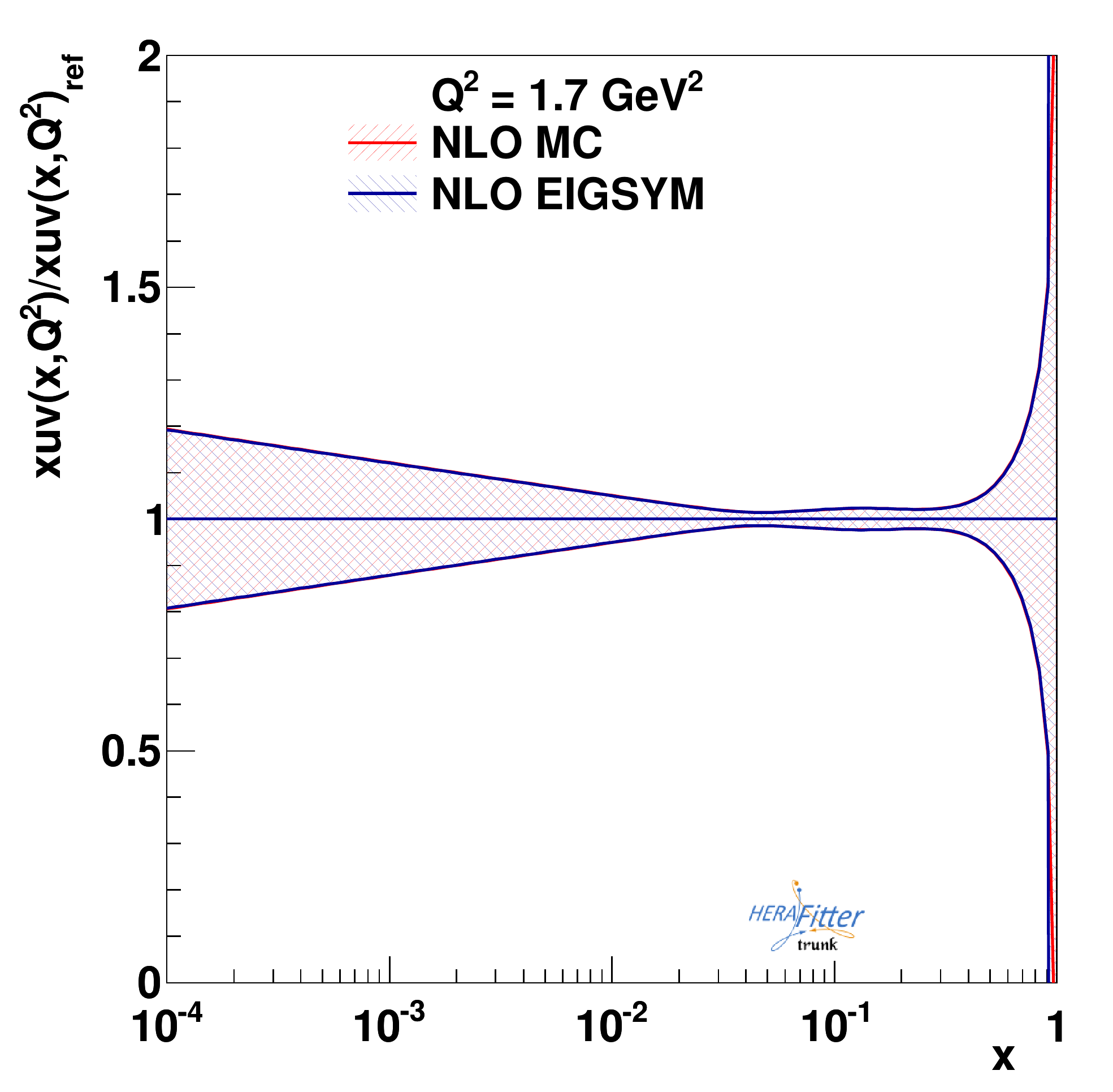} 
\end{tabular}
\begin{tabular}{cc}
\includegraphics[width=0.5\linewidth]{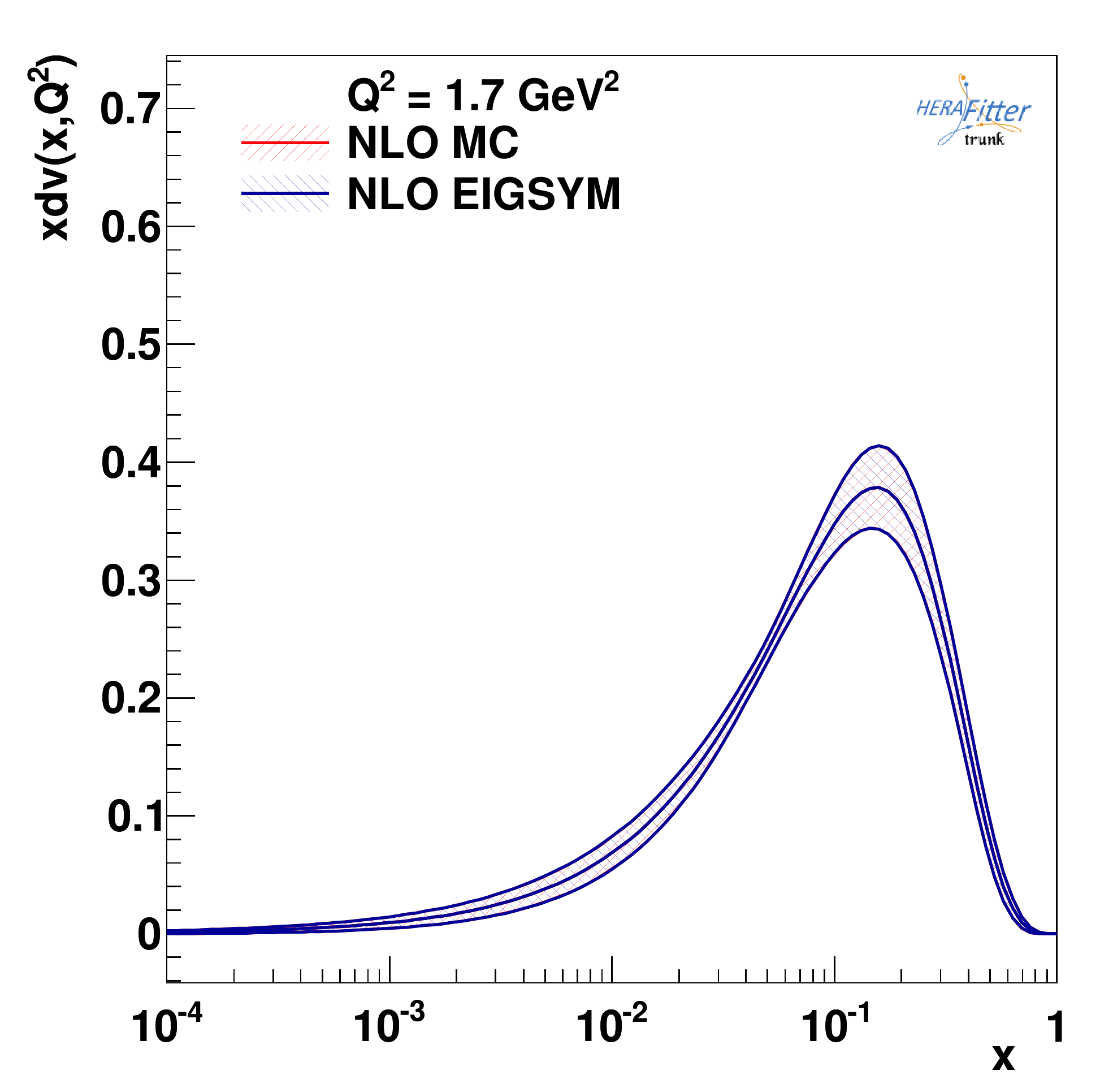} &
\includegraphics[width=0.5\linewidth]{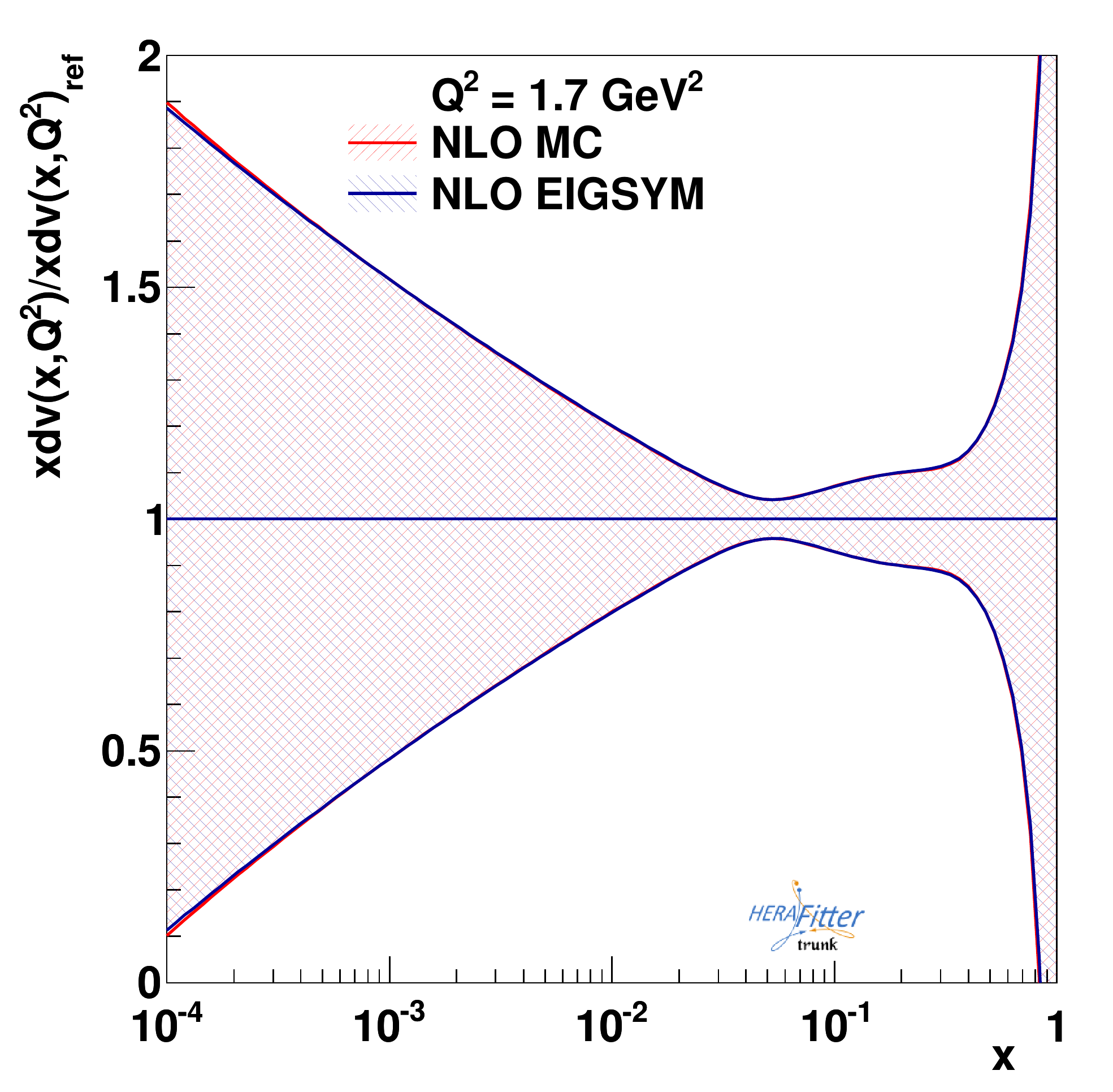} 
\end{tabular}
\begin{tabular}{cc}
\includegraphics[width=0.5\linewidth]{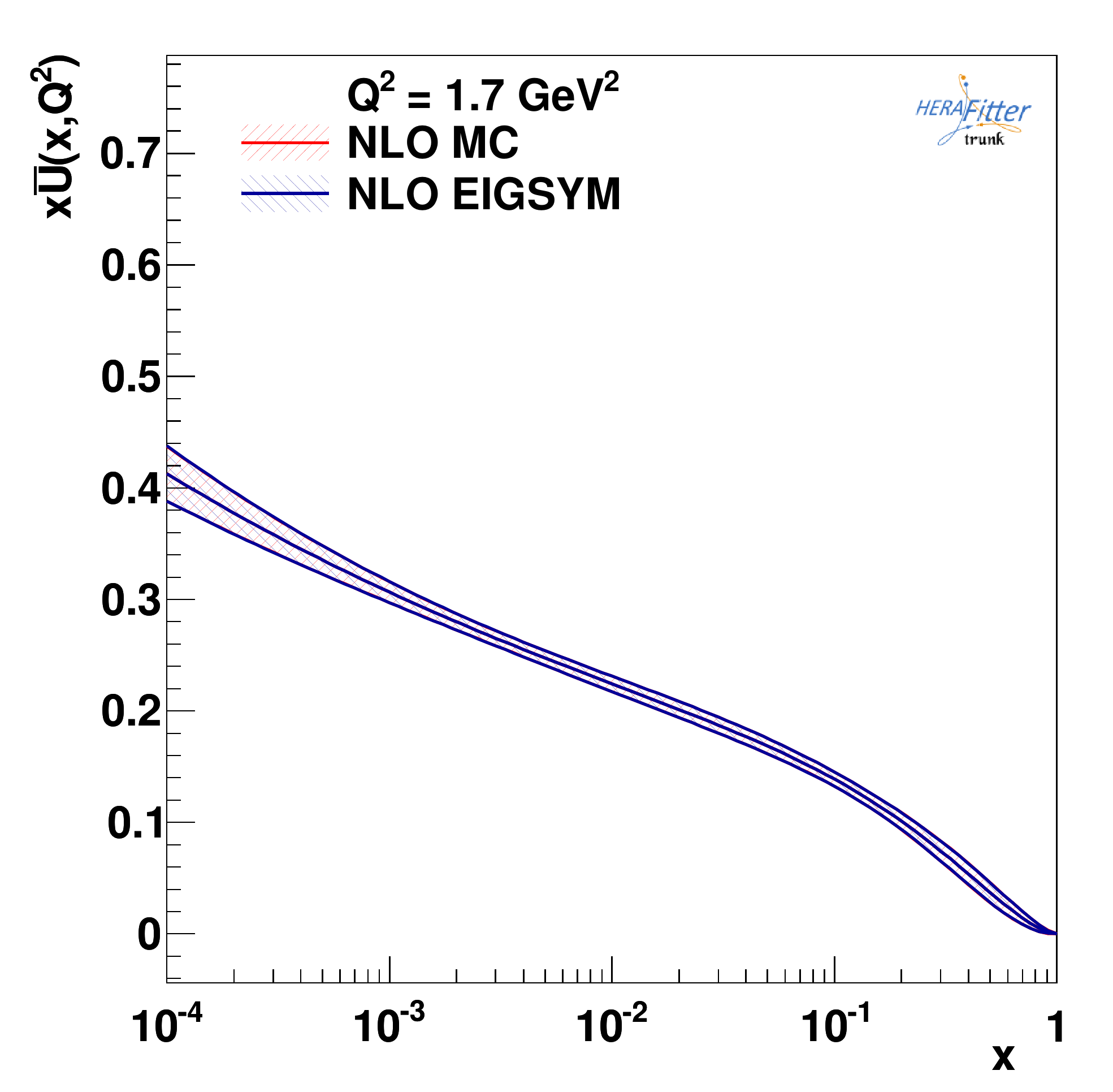} &
\includegraphics[width=0.5\linewidth]{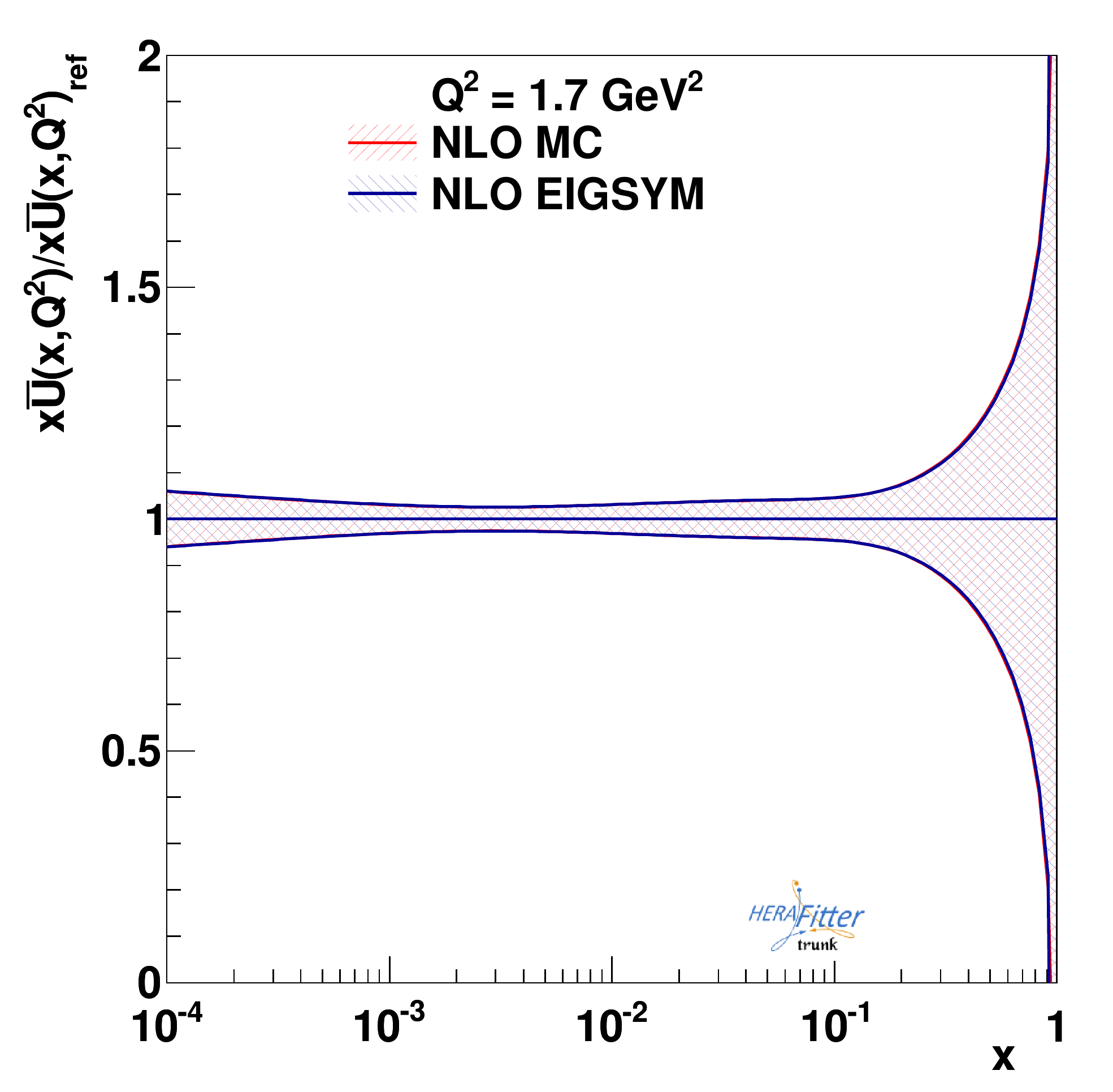} 
\end{tabular}
\begin{tabular}{cc}
\includegraphics[width=0.5\linewidth]{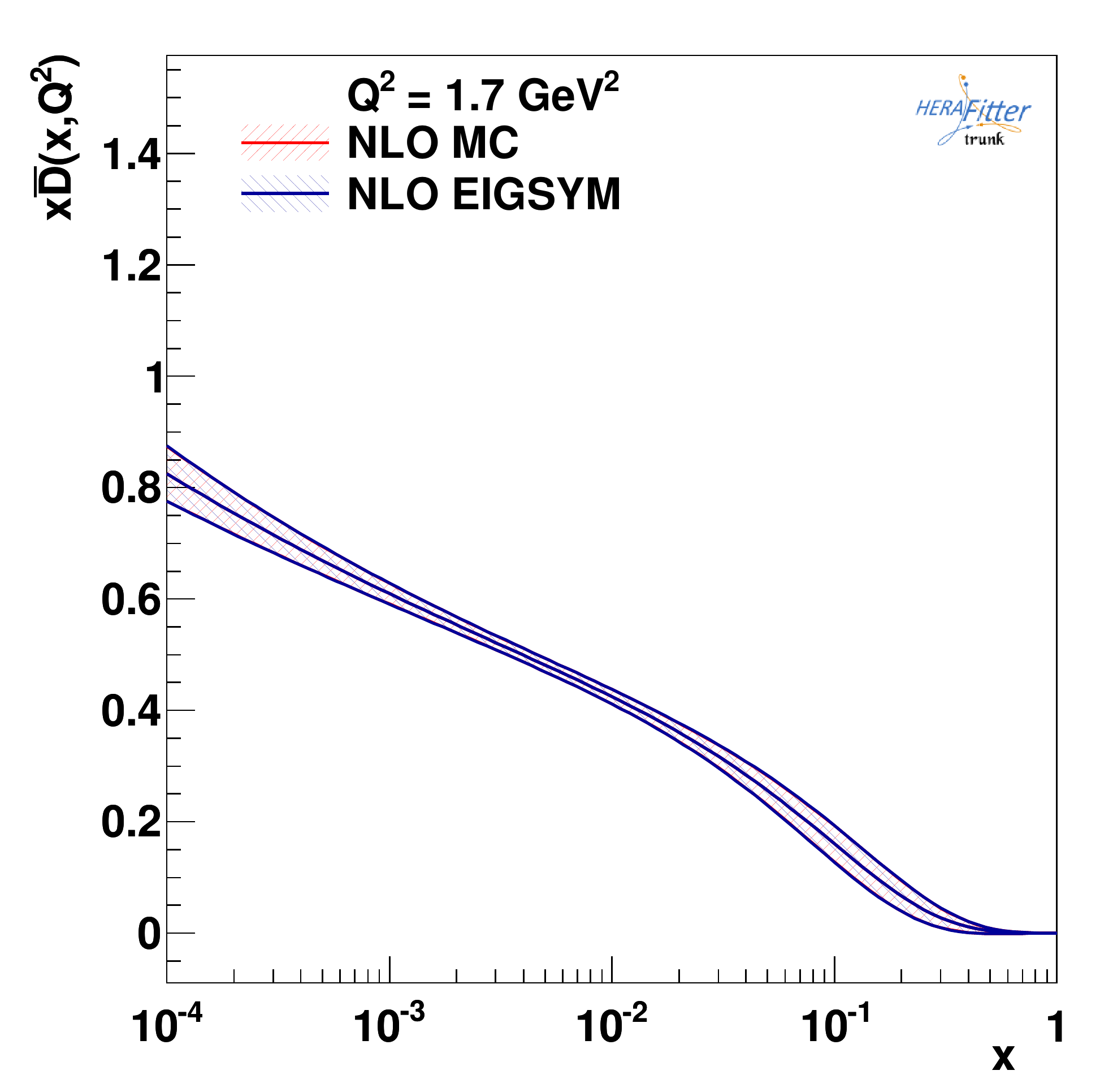} &
\includegraphics[width=0.5\linewidth]{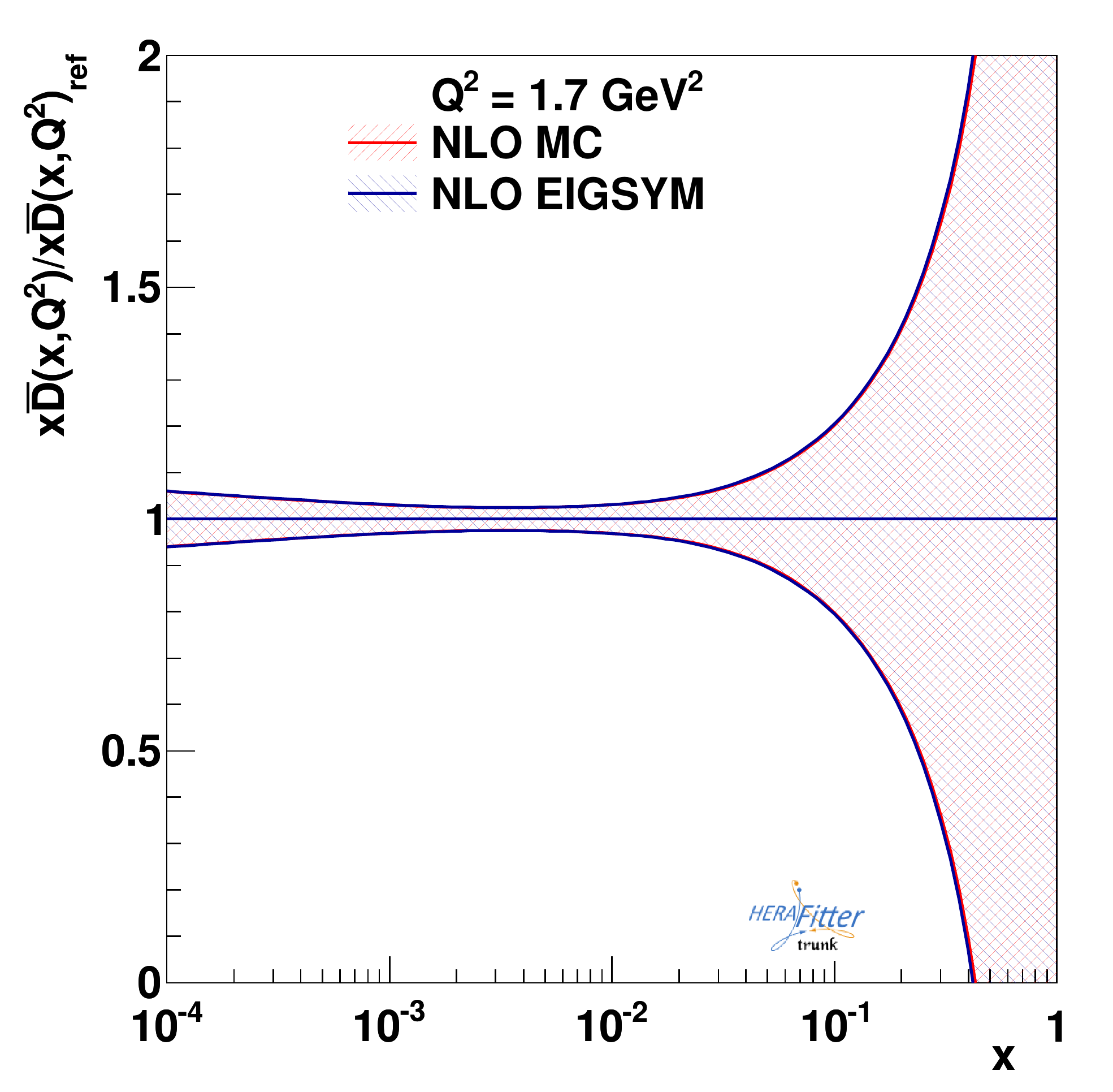} 
\end{tabular}
\begin{tabular}{cc}
\includegraphics[width=0.5\linewidth]{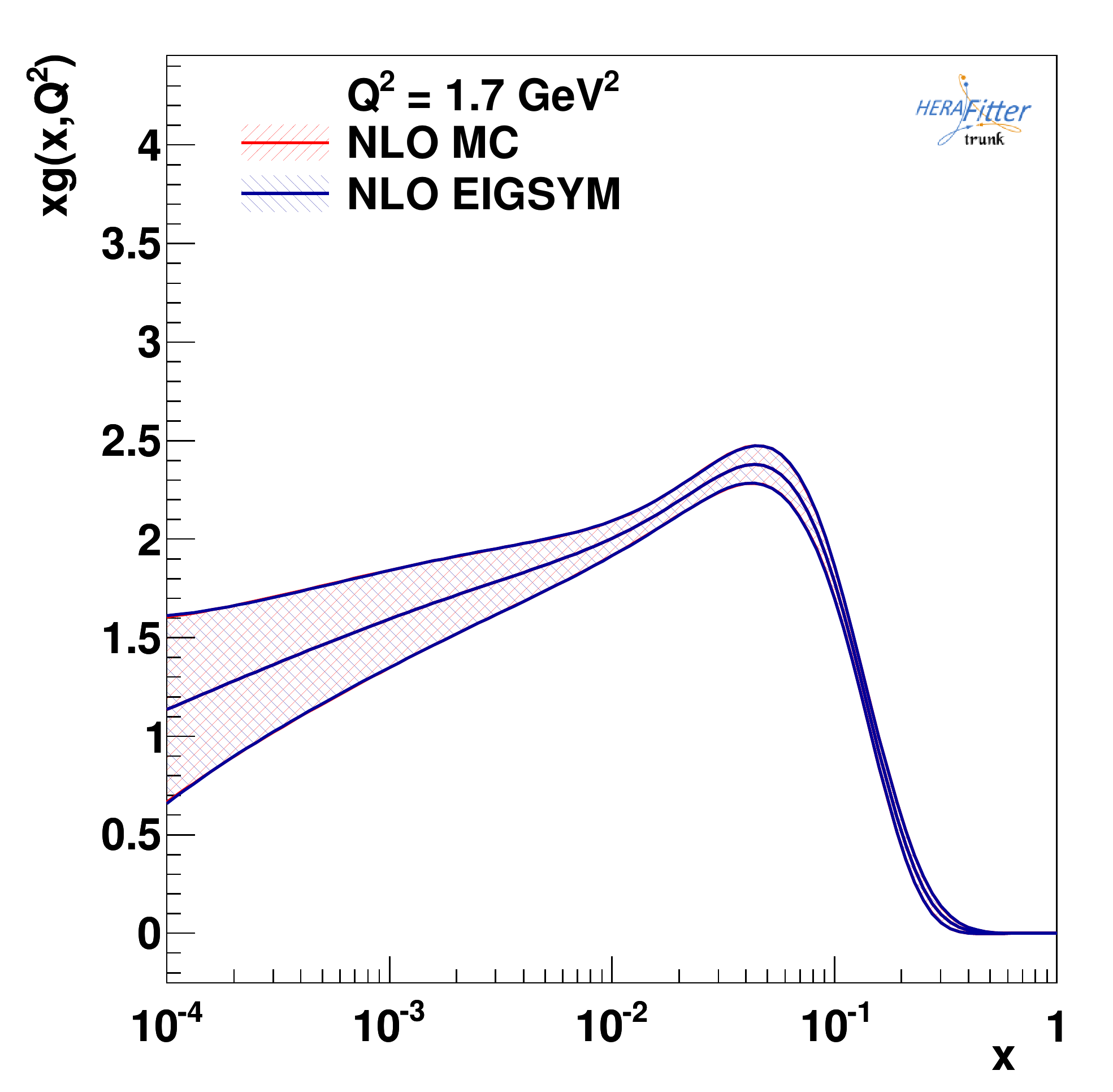} &
\includegraphics[width=0.5\linewidth]{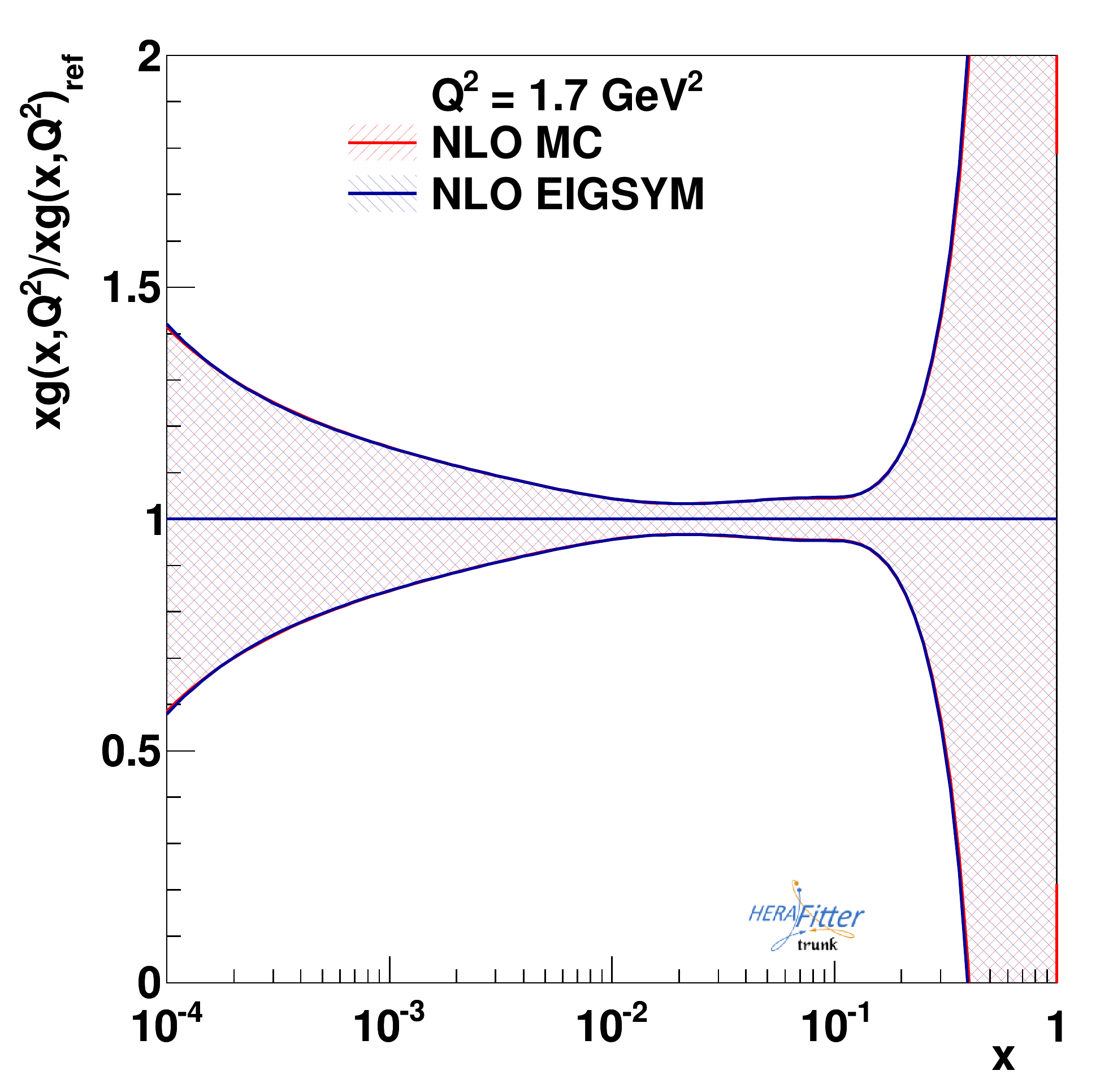} 
\end{tabular}
\end{minipage}
\end{center}
\caption{NLO PDFs with the experimental uncertainty bands as well as 
the relative uncertainties determined
by the MC method and its eigenvector representation. From top to 
bottom, the panels
show  $xu_v$, $xd_v$, $x\bar{U}$, $x\bar{D}$
and $xg$ distributions. \label{fig:PDFs}
}
\end{figure}

\begin{figure}
\begin{center}
\includegraphics[width=0.95\linewidth]{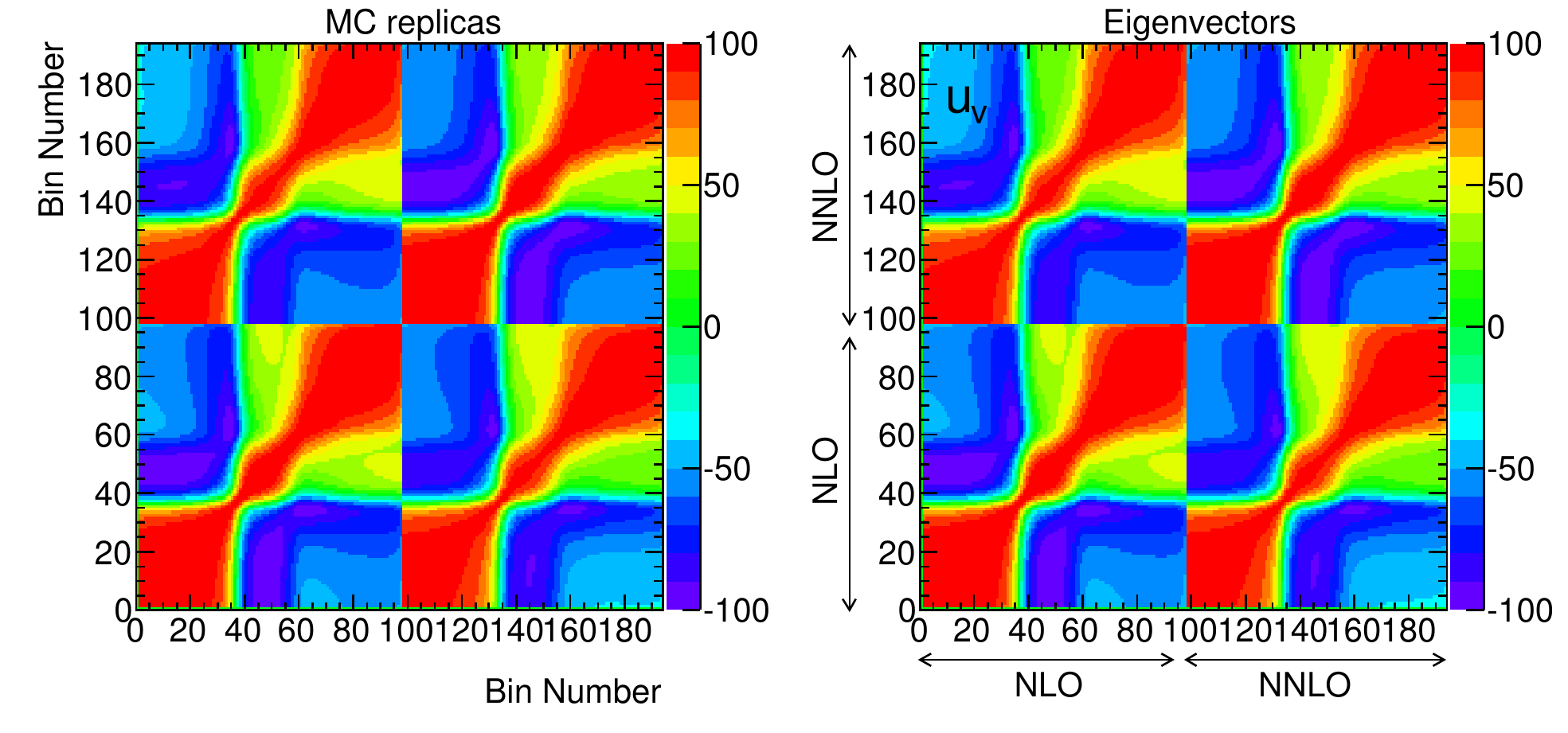}\\
\includegraphics[width=0.95\linewidth]{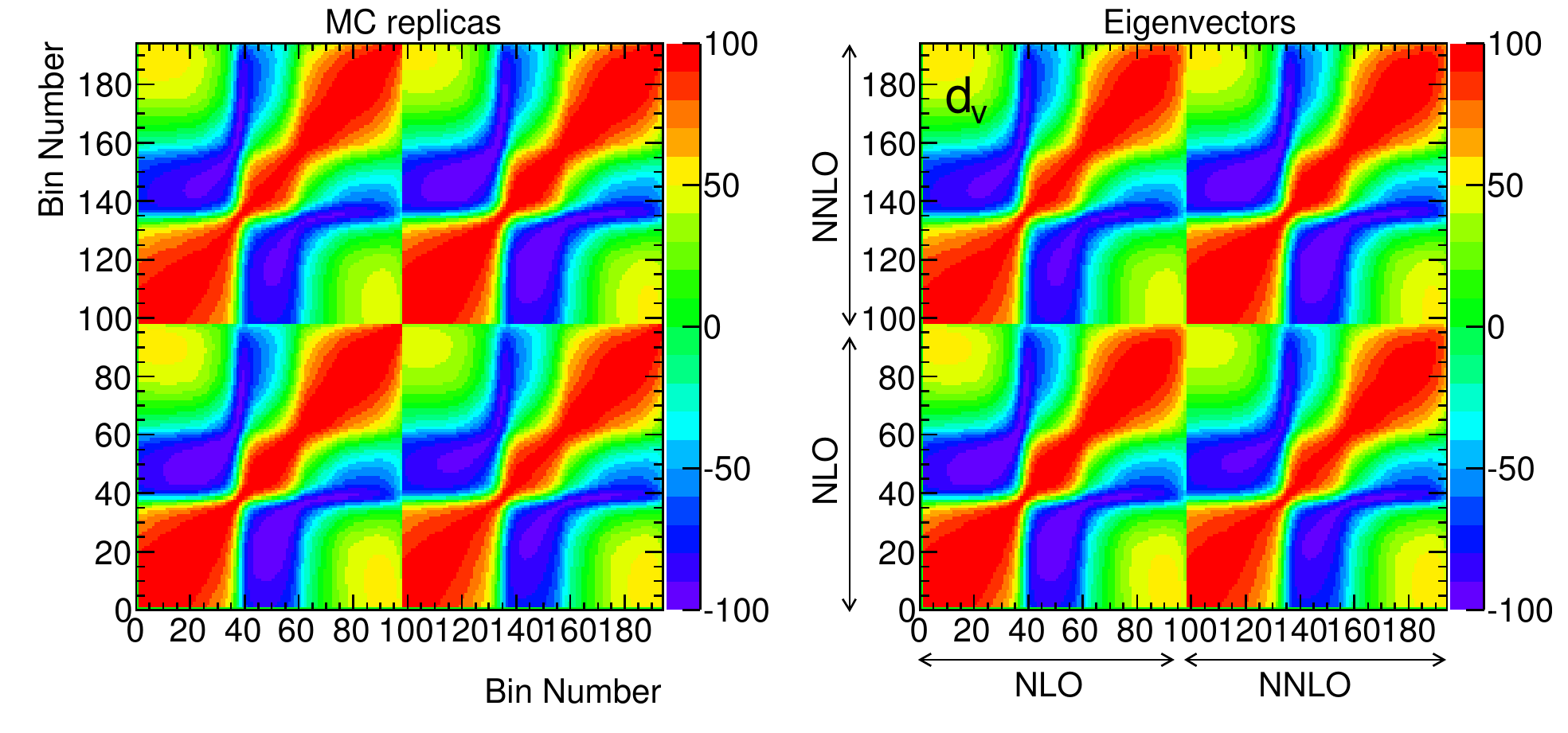}\\
\includegraphics[width=0.95\linewidth]{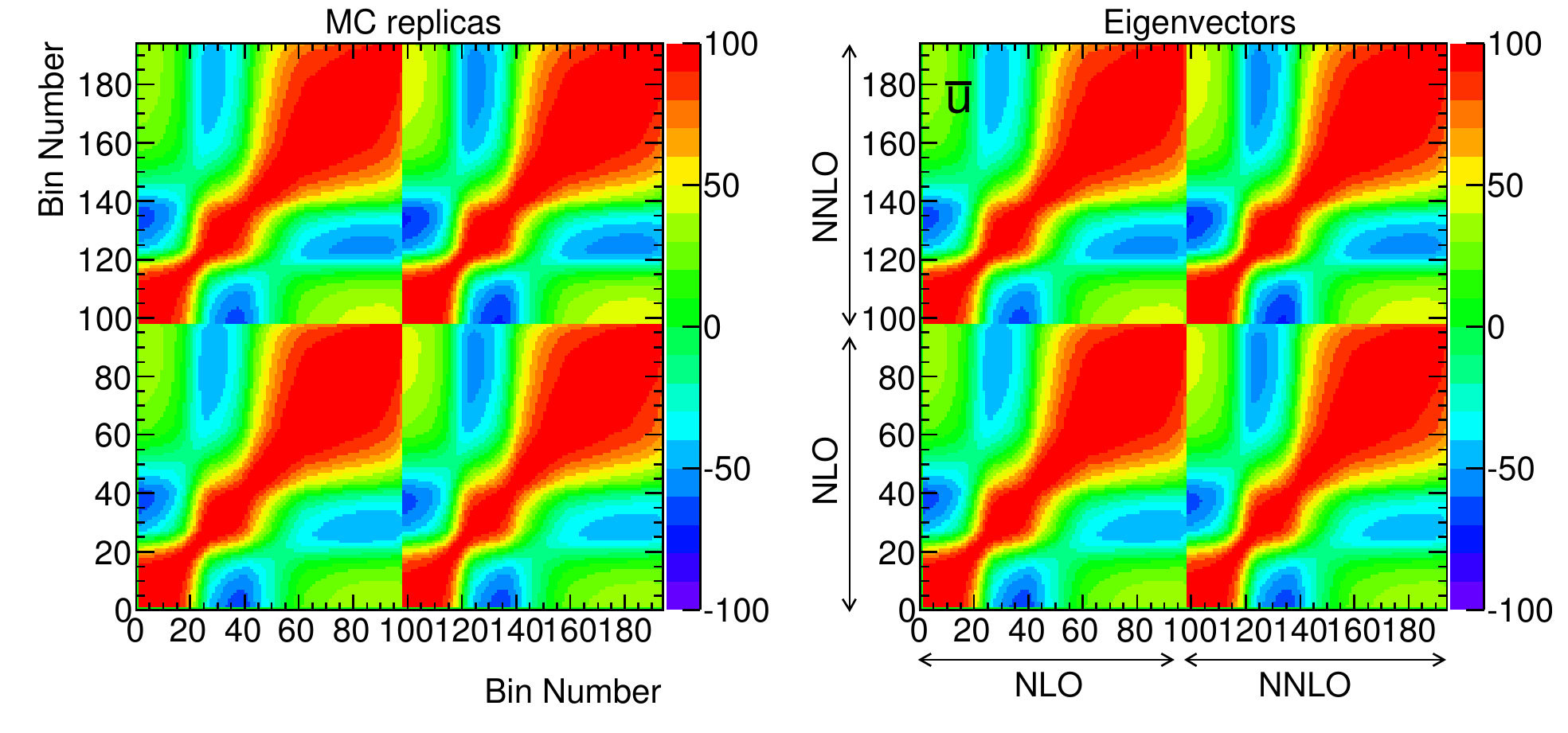}\\
\includegraphics[width=0.95\linewidth]{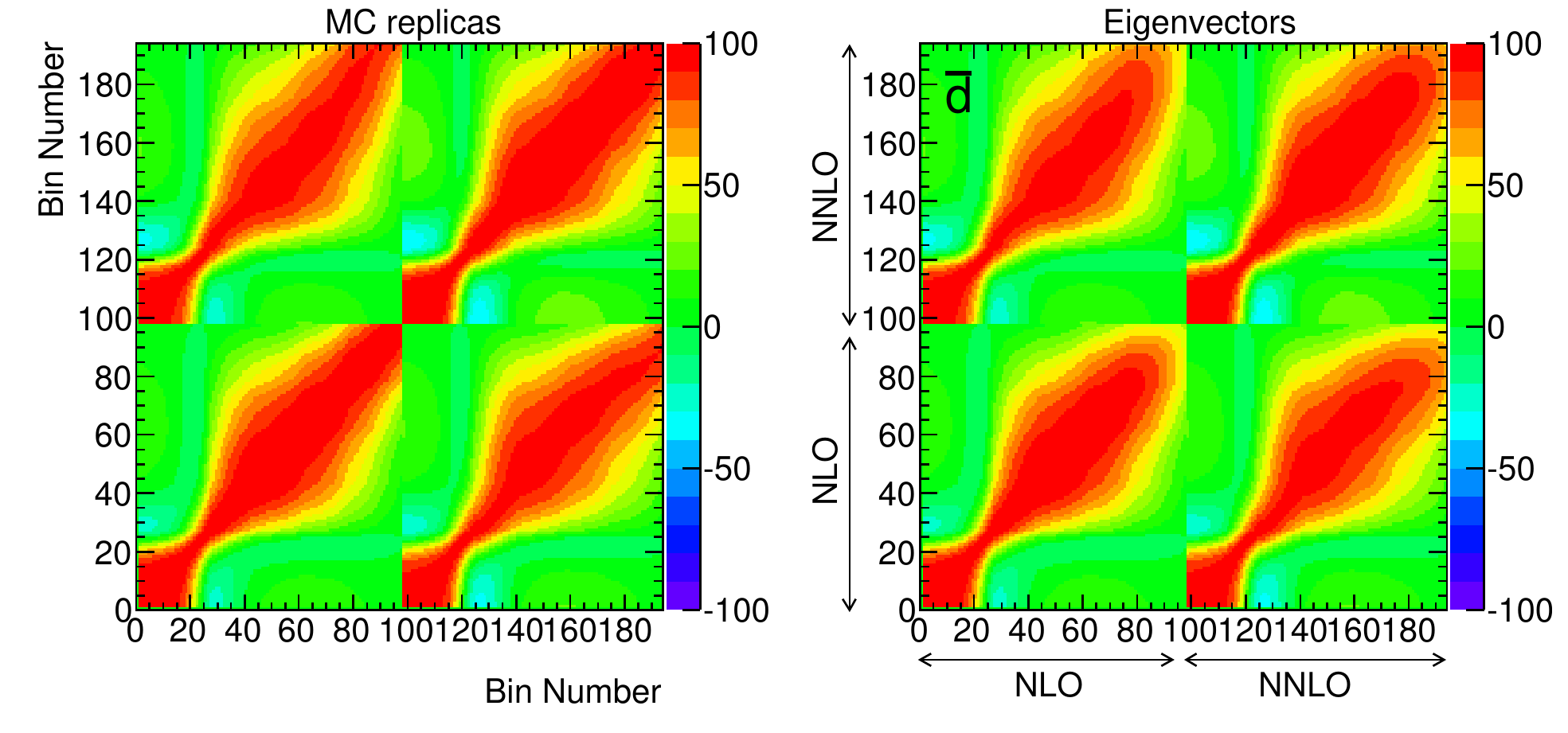}\\
\includegraphics[width=0.95\linewidth]{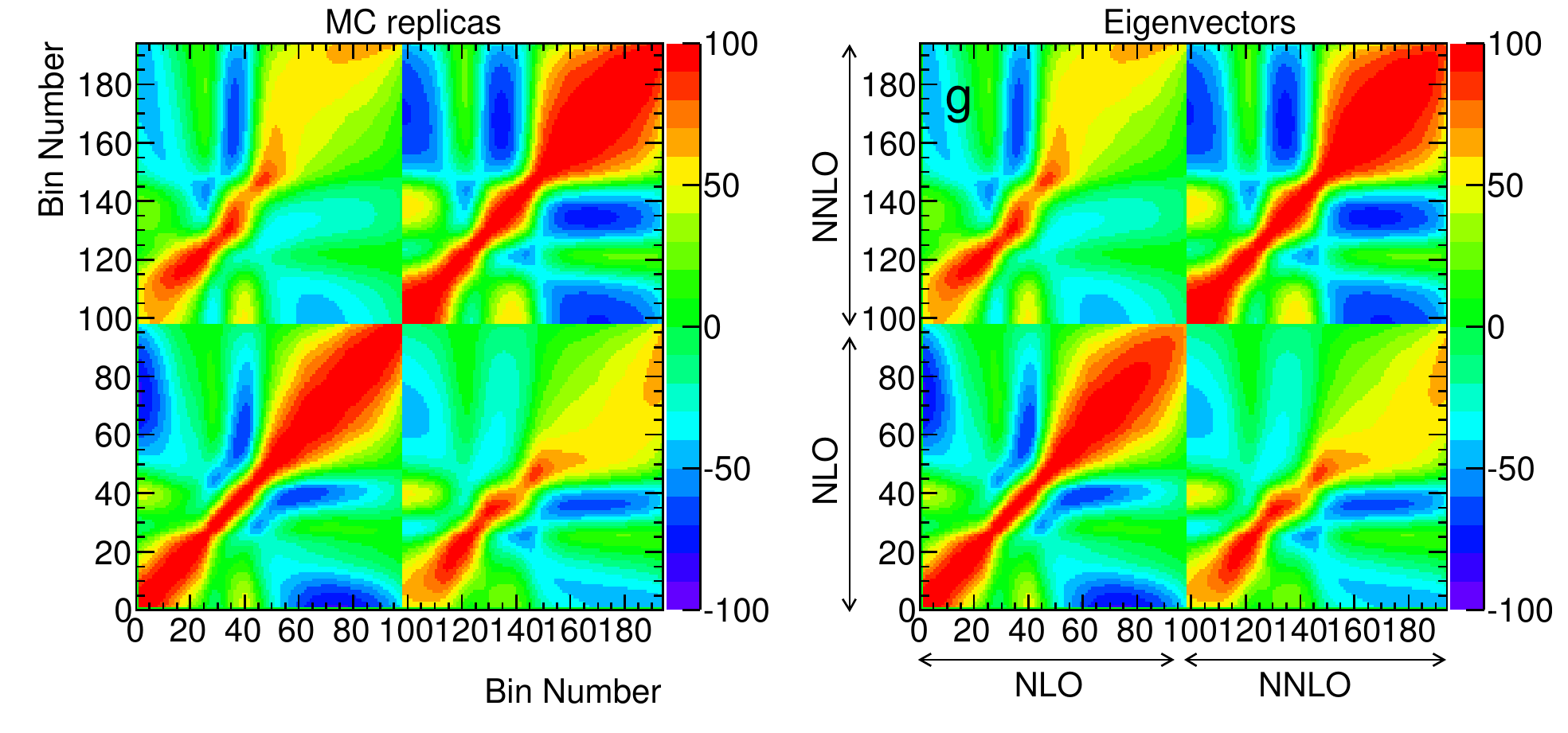}
\end{center}
\caption{Correlation coefficients, given in percent and represented 
by different colours, among different PDFs
at NLO and NNLO at the starting scale $Q^2=1.7$~GeV$^2$ 
and $x$-grid points. From top to bottom, the panels
show correlation coefficients for the $xu_v$, $xd_v$, $x\bar{U}$, $x\bar{D}$
and $xg$ distributions. The left column corresponds to the original MC-method 
calculation and the right shows the result of the eigenvector
representation. Each panel shows the correlation coefficients as a function
of the $x$-grid point for the NLO (bins $1-97$) and NNLO (bins $98-194$) PDFs. 
Bins $0$, $27$, $43$, $62$, $78$ and $97$
 correspond to anchor points at $x=10^{-5}$, $0.01$,
$0.1$, $0.4$, $0.7$ and $1.0$ with logarithmic $x$ spacing  between them.          
\label{fig:PDFcor}}
\end{figure}
For many applications, the eigenvector representation of the 
PDF uncertainties~\cite{Pumplin:2000vx,Pumplin:2001ct} is more convenient 
than the MC representation. The eigenvector 
representation typically requires fewer PDF
sets to describe the PDF uncertainties.
A procedure suggested in~\cite{Gao:2013bia} is adapted here to determine the 
eigenvector representation for
the correlated LO and  NLO as well as NLO and NNLO MC PDF sets. 

The procedure makes use of the ability of the QCDNUM program to perform PDF evolution 
based on a tabulated input.
An $x$-grid of $N_x = 97$ points $x_l$ with variable 
spacing\footnote{The grid for the central fit uses $199$ grid points 
spanning in $x$ from $10^{-6}$ to $1$ with four anchor points at $0.01$,
$0.1$, $0.4$ and $0.7$ and logarithmic spacing between them.
The grid for the error determination spans in $x$ from $10^{-5}$ to 
$1$ with the same anchor points. The uncertainties for $x<10^{-5}$ 
are set to those at $x=10^{-5}$.} is used to determine the $N_f=5$ average PDFs $x\overline{f(x_l)}$.
The PDFs are represented by  Eq.~\ref{eq:1}-\ref{eq:5} 
including correlations between PDFs at the $N_o=2$ orders, LO-NLO and NLO-NNLO. 
The correlated uncertainties are described by the dimension $N=N_x \times N_f \times N_o = 97\times 5\times 2$ 
covariance matrix $C$ which is represented as
$$
  C_{ij}  = \sum_{k=1}^N V_{ik}V_{jk}\,,
$$
where the matrix $V$ is built using eigenvectors of $C$ times the square root of the corresponding eigenvalues. For each vector $V_k$, a symmetric PDF error 
set is defined at 
the  starting scale as
$$
  x f^k(x_l) =  x \overline{f(x_l)}  + V_{ik}\,.
$$
Here the index $i$ is determined by the
$x$-grid index $l$,  PDF flavour index $f$ and  order index $o$ as 
$i = l + (f-1)N_x + (o-1)N_xN_f$. The resulting error sets 
are  evolved from the starting scale to other scales using QCDNUM.
Since the eigenvalues are found to be strongly ordered in magnitude, 
only $39$ ($45$) eigenvectors corresponding to leading eigenvalues can 
approximate the matrix $C$ for  
NLO-NNLO (LO-NLO) sets with  high precision, as demonstrated in the following discussion. 

The NLO PDFs with their uncertainties  determined using the MC method and
its eigenvector representation, using $39$ sets, are shown in Fig.~\ref{fig:PDFs}. Very
good agreement is observed between the two representations.
A similar picture is observed for the LO and  NNLO PDFs.
The correlation among PDF values at different $x$  is shown in Fig.~\ref{fig:PDFcor}. 
The eigenvector 
representation reproduces all the correlations very well with
small  deviations at high $x$ ($x>0.7$).
All PDFs show high degree of  correlation for neighbouring 
$x$ values which can be explained by intrinsic smoothness of the
PDF parameterisation, which has few parameters, and the fact that the 
PDFs at comparable $x$ are constrained by similar input data. 
There is a sizeable anti-correlation 
between PDFs at small and large $x$ values caused by sum rules. 
The correlation patterns as a function of $x$
 are similar for PDFs determined at NLO and NNLO 
and, with the exception of the gluon density at high $x$, there is a strong correlation between NLO and NNLO PDFs.
A qualitatively similar, strong correlation is 
observed for the PDFs determined at LO and NLO; however, it is somewhat reduced compared
to that for the NLO and NNLO PDFs. This explains why more eigenvectors are required for
the correlated LO-NLO PDF set. As a cross check, the correlations
between NLO and NNLO PDFs are studied using a bi-log-normal parameterisation
$$
 x f(x) =a x^{p - b\log(x)} (1-x)^{q-d \log(1-x)} 
$$
 instead
of the parameterisation of Eq.~\ref{eq:1}-\ref{eq:5}. Similar correlation patterns
are observed with some differences for the gluon density at high $x$, where the
uncertainties are large.

Model uncertainties in PDFs arise from the uncertainties 
of the input parameters of the fit. 
The value of the strange-quark density suppression
$r_s$ is varied by $\pm 0.30$. The variation range is defined 
by the uncertainties found by the ATLAS collaboration~\cite{Aad:2012sb,Aad:2014xca}
and cover the somewhat lower value determined by the CMS collaboration~\cite{Chatrchyan:2013mza,Chatrchyan:2013uja}.
Based on the ATLAS analysis, this variation is 
considered to be fully correlated between the NLO and NNLO PDFs.  

The uncertainties of the heavy-quark masses are also assumed to be fully 
correlated between NLO and NNLO. 
The charm-quark mass uncertainty is taken
from the H1 and ZEUS analysis~\cite{Abramowicz:1900rp} to be $0.06$ GeV.  
The bottom-quark mass is varied between $4.3$ and $5.0$ GeV.

The uncertainties of the QCD evolution at small $Q^2$ are probed by
varying the $Q^2_{\rm min}$ cut between $5$ and $10$~GeV$^2$. 
The choice of the $Q^2_0$ value is also tested by varying  down
to $Q^2_0 = 1.5$~GeV$^2$. The resulting change in the PDFs is considered as a symmetric
uncertainty.

The strong coupling constant at both NLO and NNLO, may be considered to be the 
same, or different, following the analyses from~\cite{Lai:2010vv,Aaron:2009aa} or~\cite{Alekhin:2013nda,mstw2008}, respectively. To cover different possibilities, 
$\alpha_S(M_Z)$ is varied by $\pm 0.002$ independently for the LO, NLO and NNLO fits.

Parameterisation uncertainties are estimated by including additional
terms in the polynomial expansion following the procedure outlined 
in~\cite{Aaron:2009aa}. The extra terms are added coherently to
LO, NLO and NNLO sets to preserve the correlation pattern.

The PDF sets are reported 
in the  LHAPDF v6 format~\cite{lhapdf}. 
The correlated NLO-NNLO and LO-NLO sets are labelled as ``HF14cor-nlo-nnlo'' and ``HF14cor-lo-nlo'', 
respectively. Separate
sets are provided for experimental and model plus parameterisation (``HF14cor-lo-nlo-nnlo\_VAR'')
uncertainties. The experimental uncertainties are reported as both
Monte Carlo (``HF14cor-lo-nlo-nnlo\_MC'') and symmetric eigenvector (``HF14cor-nlo-nnlo\_EIGSYM''
, ''HF14cor-lo-nlo\_EIGSYM'') sets. The symmetric eigenvector set is ordered according to the 
size of the PDF uncertainty, approximate calculations may use the first $26$ sets only. 
The reference set for all PDF sets
is chosen to be the set averaged  over the MC replicas.

\section{Prediction of $Z$ and $WW$ production cross sections at the LHC}
\begin{figure}
\begin{center}
\includegraphics[width=0.85\linewidth]{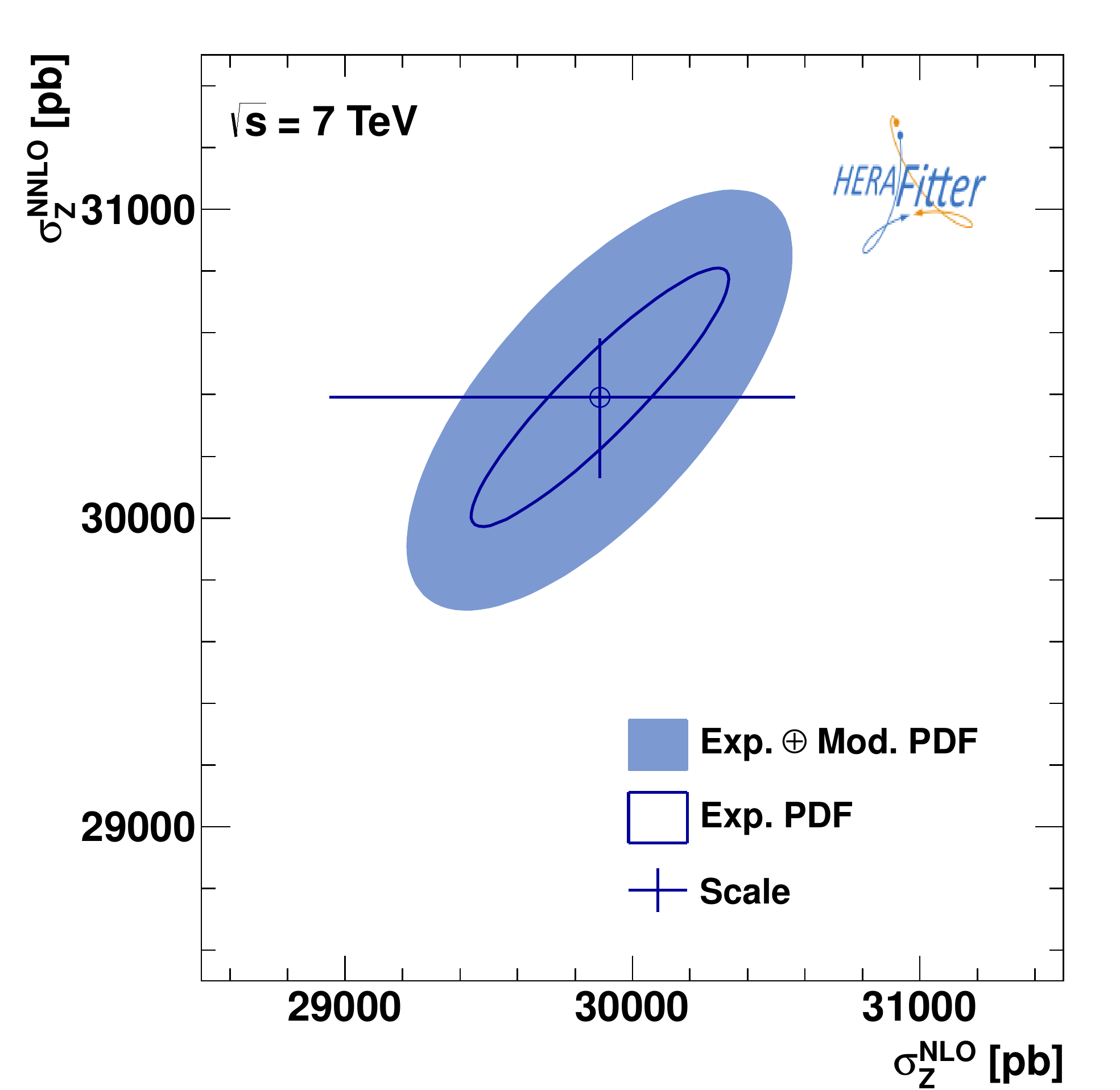}\\
\includegraphics[width=0.85\linewidth]{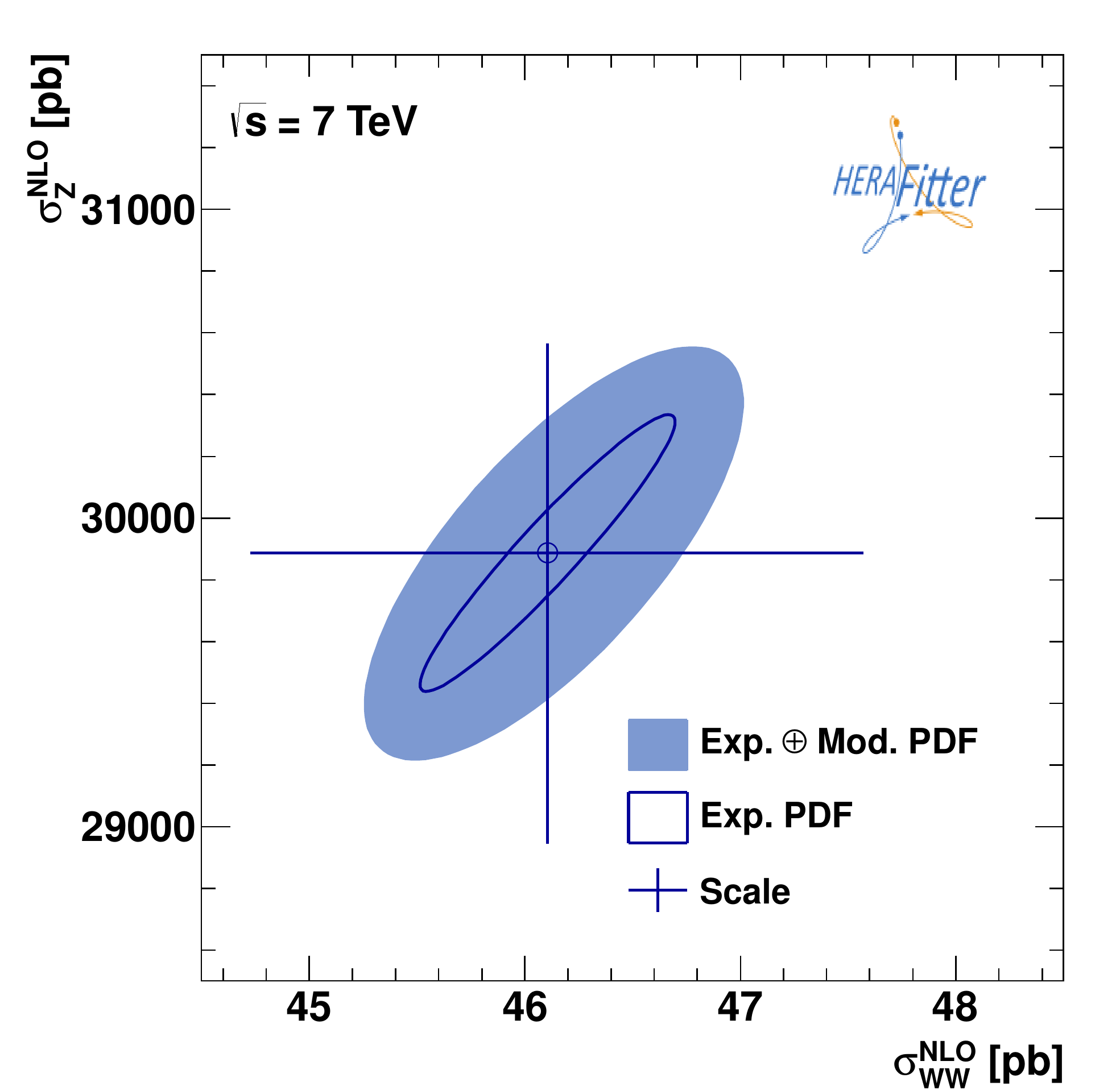}\\
\includegraphics[width=0.85\linewidth]{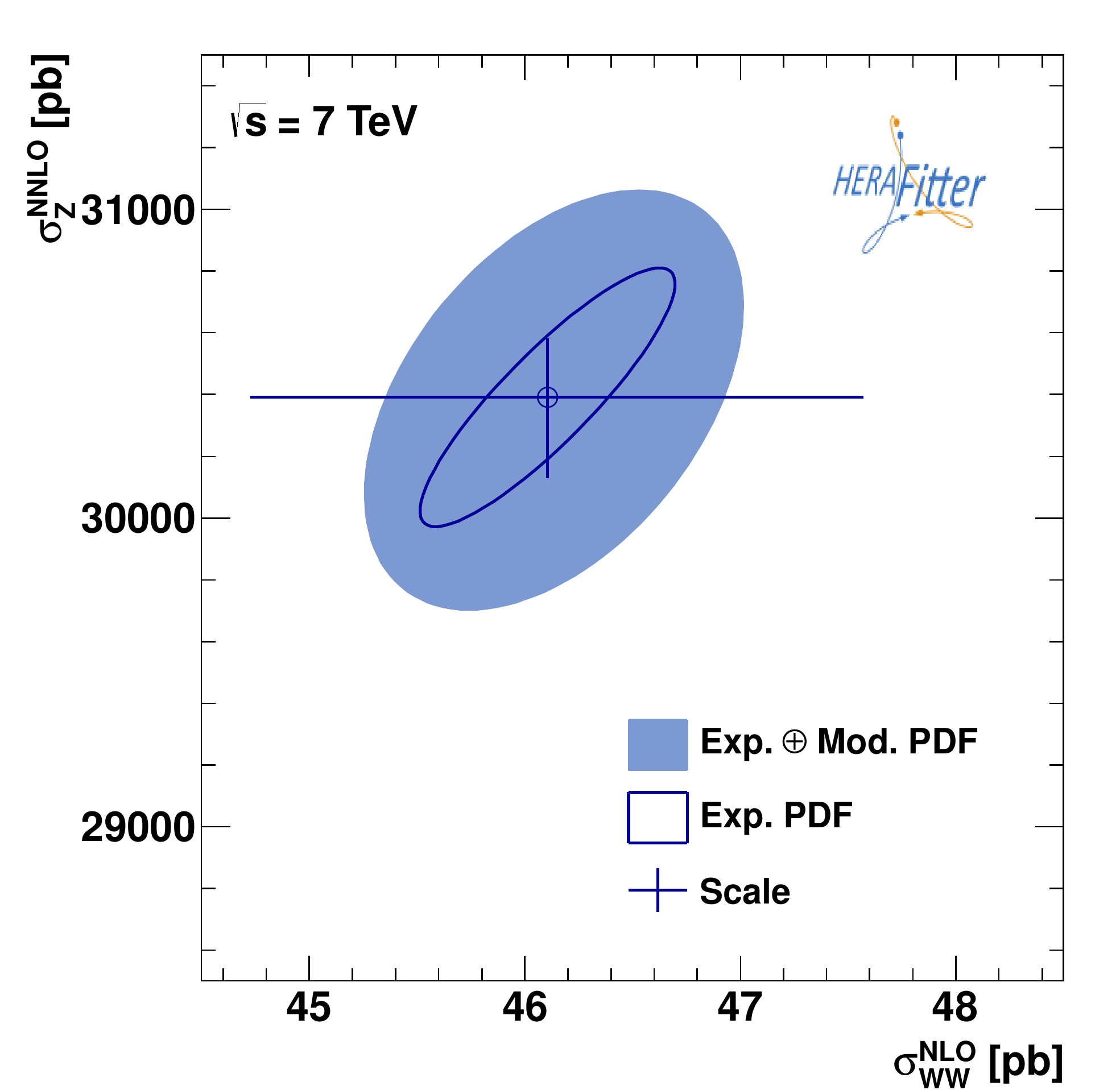}
\end{center}
\caption{\label{fig:corrwwz}Correlation of the cross-section 
predictions for $Z$ boson production calculated at NLO and NNLO,
$WW$ di-boson and $Z$ boson production both calculated at NLO and
 $WW$ di-boson production calculated at NLO and $Z$ boson production
calculated at NNLO. 
The error bars indicate  scale uncertainties.}  
\end{figure}
\begin{table*}
\begin{center}
\renewcommand{\arraystretch}{1.25}
\begin{tabular}{c|ccccccc}
\hline
Cross section     & Value  & Exp. PDF & Mod. PDF & Scale  &  \multicolumn{3}{c}{Correlation coefficient} \\
                  & pb     & pb       & pb        & pb     &    $\sigma_Z^{\rm NLO}$ & $\sigma_Z^{\rm NNLO}$  & $\sigma_{WW}^{\rm NLO}$ \\
\hline
$\sigma_Z^{\rm NLO}$    & 29890 & $\pm450$ & ${}^{+490}_{-490}$ & ${}^{+680}_{-940}$  & 1 & 0.697 & 0.736\\
$\sigma_Z^{\rm NNLO}$   & 30390 & $\pm420$ & ${}^{+520}_{-540}$ & ${}^{+190}_{-260}$   & 0.697 & 1 & 0.451\\
$\sigma_{WW}^{\rm NLO}$ & 46.1  & $\pm0.6$ & ${}^{+0.7}_{-0.6}$ & ${}^{+1.5}_{-1.4}$  & 0.736 & 0.451 & 1\\
\hline 
\end{tabular}
\caption{\label{tab:tabs} Cross-section predictions, experimental (Exp.) as well as model and parameterisation (Mod.) PDF uncertainties, scale uncertainties and correlation
coefficients for the $Z$ boson and $WW$ di-boson production calculated at NLO and NNLO using the 
HF14cor-nlo-nnlo PDF set.}
\end{center}
\end{table*}
The usage of the correlated NLO and NNLO PDF sets is exemplified by
calculating
$WW$ di-boson and $Z$ boson production cross sections 
for the $pp$ collisions at a $\sqrt{s}=7$~TeV centre-of-mass energy.
The recent measurements  of $WW$ di-boson production by the ATLAS
and CMS collaborations~\cite{ATLAS:2012mec,Chatrchyan:2013yaa} have generated considerable
interest from the theoretical community. The uncertainties of the measurements and predictions are comparable and the
measurements are about $1-2\sigma$ above the expectations. The difference
may originate from missing higher orders~\cite{Dawson:2013lya,Campanario:2013wta}, electroweak 
effects \cite{Billoni:2013aba}
and possible New Physics contributions~\cite{Rolbiecki:2013fia}.   

The $WW$ di-boson and $Z$ boson production
processes are expected to have similar PDF dependences which 
may lead to reduced uncertainties for the ratio of the cross sections.
In the following discussion, the predictions obtained using the HF14cor-nlo-nnlo PDF sets
 are compared to the measurement of the ratio
 obtained by the CMS collaboration~\cite{Chatrchyan:2013yaa}.

The total cross section for $W^+W^-$ di-boson production, $\sigma_{WW}$, 
(called $WW$ di-boson production in the following)
is calculated at NLO
using the MCFM v6.6 program~\cite{Campbell:2011bn,Campbell:2011cu}.
The calculation includes the gluon-gluon
initiated box diagram which first contributes at order $\alpha^2_S$ and
so is formally NNLO.
The factorisation and renormalisation scales are given by half of 
 the scalar sum of the transverse momenta of the outgoing final-state particles, $H_T/2$.   
The contribution from Higgs boson production, which contributes
approximately two percent, is not included.
As a cross check, the total $WW$ di-boson cross-section predictions from the original 
paper~\cite{Campbell:2011cu} are reproduced using the corresponding setup.

The total cross section for $Z/\gamma^*$ boson production, $\sigma_Z$,
(referred to as  $Z$ boson production in the following discussion) is calculated at NLO 
and  NNLO using FEWZ~\cite{fewz,Li:2012wna}. 
The invariant mass for the lepton pair is chosen to be $60<M_{\ell\ell}<120$~GeV as in the  analysis
of the CMS collaboration. 
The factorisation and renormalisation scales are fixed to the $Z$ boson
pole mass, $M_Z$. 
The FEWZ calculation includes NLO electroweak corrections,
which are small for this mass range. The contribution from
$\gamma\gamma \to \ell \ell$ processes is not included for either $WW$ di-boson or  
$Z$ boson production. 

Uncertainties due to missing higher-order corrections
are estimated by varying the default scale up and down by a factor of two,
for both factorisation and renormalisation scales simultaneously or independently, 
excluding the variation in  opposite directions. An envelope
of all variations is built and maximal positive and negative deviations are taken as the asymmetric uncertainty. 
The scale uncertainty is dominated by the variation of the renormalisation scale for $WW$ di-boson production and by the variation of the factorisation scale for $Z$ boson production.
The scale uncertainty is treated as uncorrelated
between $WW$ di-boson and $Z$ boson production.
The experimental PDF uncertainties are symmetric
by construction. The model and parameterisation PDF uncertainties are quoted as asymmetric.

The resulting cross sections with their correlations are given in
Table~\ref{tab:tabs} and shown in Fig.~\ref{fig:corrwwz}. The
predictions for $Z$ boson production calculated at NLO and NNLO  show a
high degree of correlation. 
The scale uncertainties are reduced significantly for the NNLO prediction, becoming  smaller than the 
 PDF uncertainties.
The central value of the  prediction at NNLO
is larger than that for NLO by $1.7\%$. This difference is smaller than the
uncertainty of $\sigma_Z^{\rm NLO}$ on the missing higher order corrections,
estimated by the scale variation.

The correlation of the $\sigma_{WW}$ and $\sigma_Z$ cross sections is very
large for the experimental PDF uncertainties for both the NLO and NNLO calculations. 
Model and parameterisation PDF uncertainties  are also highly correlated for most
of the uncertainty sources  when both cross sections are calculated at NLO. 
When $\sigma_Z$ is calculated
at NNLO, an anti-correlation for some sources is observed. A detailed
breakdown of the model and parameterisation uncertainties for the total cross-section calculations is given in
Table~\ref{tab:tabmodel}.  
\begin{table}
\begin{center}
\renewcommand{\arraystretch}{1.25}
\begin{tabular}{c|ccc}
\hline
Variation     & $\sigma_{WW}^{\rm NLO}$  & $\sigma_Z^{\rm NLO}$ & $\sigma_Z^{\rm NNLO}$ \\
              & \%      & \%     & \%     \\
\hline
$r_s (-0.3)$	&	1.00	&	-0.29	&	-0.33	\\
$r_s (+0.3)$	&	-0.81	&	0.39	&	0.42	\\
$M_c (-0.06$~GeV$)$	&	-0.81	&	-0.89	&	-0.76	\\
$M_c (+0.06$~GeV$)$	&	0.55	&	0.66	&	0.61	\\
$M_b (-0.45$~GeV$)$	&	0.13	&	0.11	&	-0.02	\\
$M_b (+0.25$~GeV$)$	&	-0.07	&	-0.07	&	0.00	\\
$\alpha_S(M_Z) (-0.002)$	&	-0.54	&	-1.27	&	-1.17	\\
$\alpha_S(M_Z) (+0.002)$	&	0.52	&	1.23	&	1.17	\\
$Q^2_{\rm min} (-2.5$~GeV$^2)$	&	-0.25	&	-0.35	&	0.23	\\
$Q^2_{\rm min} (+2.5$~GeV$^2)$	&	0.75	&	0.73	&	-1.06	\\
$Q^2_0 (-0.2$~GeV$^2)$	&	-0.21	&	-0.19	&	-0.14	\\
+$D_{u_v}$	&	-0.03	&	-0.32	&	0.97	\\
+$D_{\bar{U}}$	&	-0.04	&	-0.02	&	-0.01	\\
+$E_{\bar{U}}$	&	0.01	&	0.00	&	0.00	\\
\hline
\end{tabular}
\caption{\label{tab:tabmodel}Shifts of the $WW$ di-boson and $Z$ boson production cross sections due to the model and parameterisation variations in the PDF fit.}
\end{center}
\end{table}
An anti-correlation between $\sigma_{WW}$ and $\sigma_Z$ 
is observed for the variation of the $r_s$ parameter.
In addition, an anti-correlation between  $\sigma^{\rm NLO}_{Z}$ and $\sigma^{\rm NNLO}_{Z}$ 
is observed for 
the variation of the $Q^2_{\rm min}$ cut as well from the addition of 
the $D_{u_v}$ parameter to the PDF parameterisation. 
A positive correlation between $\sigma_{WW}$ and $\sigma_{Z}$ at both orders is observed for 
the $M_c$, $M_b$ and $\alpha_S(M_Z)$ variations.

\begin{figure}[tb]
\centerline{\includegraphics[width=1.0\linewidth]{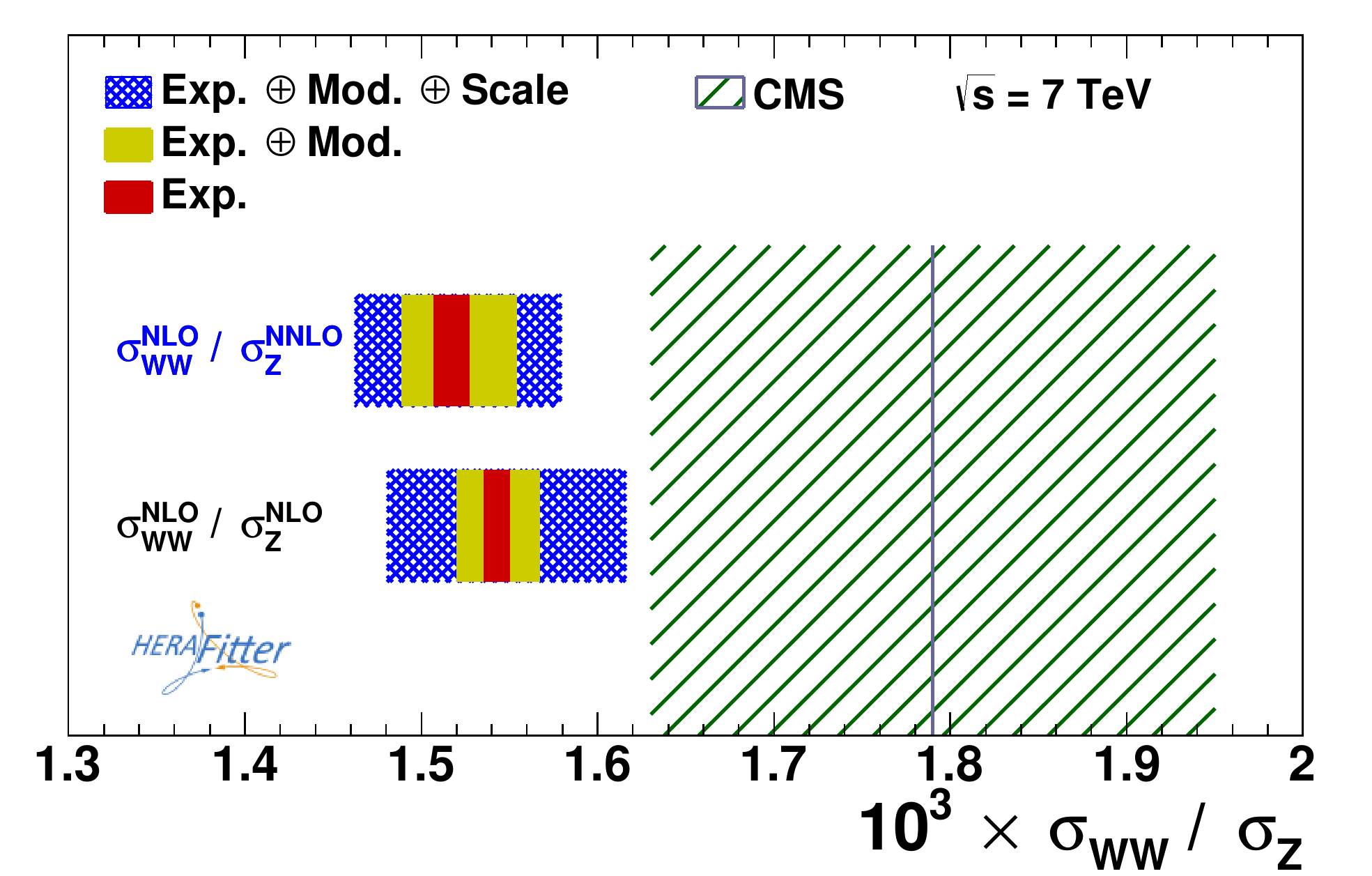}}
\caption{\label{fig:wwzratio}Ratio of the $WW$ di-boson to $Z$ boson
production cross sections calculated at NLO and NLO/NNLO compared to the
result obtained by the CMS collaboration (hatched area). The inner, middle and outer filled error bars of the
predictions indicate experimental and full PDF uncertainties and the total uncertainty calculated as the scale
and full PDF uncertainties added in quadrature, respectively.}
\end{figure}

The predicted ratio $\sigma_{WW} / \sigma_Z$ using the $Z$ boson production cross sections 
calculated at NLO and NNLO 
is given in Table~\ref{tab:tabratio}.
\begin{table}
\begin{center}
\renewcommand{\arraystretch}{1.5}
\begin{tabular}{c|cccc}
\hline
\small{Ratio}     	& \small{Value}  & \small{Exp. PDF} & \small{Mod. PDF} & \small{Scale}  \\
          	& $\times 10^{-3}$     & $\times10^{-3}$       & $\times 10^{-3}$        &  $\times 10^{-3}$    \\
\hline\\[-0.4cm]
$\frac{\textstyle \sigma_{WW}^{\rm NLO} }{\textstyle \sigma_Z^{\rm NLO}}$   & 1.543 & $\pm0.008$ & ${}^{+0.023}_{-0.021}$ & ${}^{+0.069}_{-0.058}$\\[0.2cm]
$\frac{\textstyle \sigma_{WW}^{\rm NLO}} {\textstyle \sigma_Z^{\rm NNLO}}$  & 1.517 & $\pm0.010$ & ${}^{+0.036}_{-0.027}$ & ${}^{+0.050}_{-0.046}$\\[0.2cm]
\hline
\end{tabular}
\caption{\label{tab:tabratio} Predictions of the $WW$ di-boson to $Z$ boson production cross-section 
ratio with PDF and scale uncertainties.}
\end{center}
\end{table}
The predictions are compared to the CMS data in Fig.~\ref{fig:wwzratio}. 
The data and calculations agree reasonably well. 
The scale uncertainty is  reduced by using $\sigma^{\rm NNLO}_Z$.
Experimental PDF uncertainties cancel in the ratio becoming negligible
compared to the scale uncertainties.

A detailed breakdown of the model uncertainty sour\-ces for the ratio of the cross sections
is given in Table~\ref{tab:tabratiomodel}. The $r_s$ variation results in a large uncertainty
for the ratio using both NLO and NNLO calculations of $\sigma_Z$. 
Additional experimental input constraining  $r_s$ will
allow this uncertainty to be reduced.
Variations of the $Q^2_{\rm min}$ cut and addition of the $D_{u_v}$ parameter
cancel in the ratio for $\sigma^{\rm NLO}_{WW}/\sigma^{\rm NLO}_Z$; however, these variations have significant impact
on $\sigma^{\rm NLO}_{WW}/\sigma^{\rm NNLO}_Z$.
The variations of $M_c$, $M_b$ and $\alpha_S(M_Z)$ do not affect the ratio significantly for either the  NLO or NNLO
calculations of $\sigma_Z$. 

\begin{table}
\begin{center}
\renewcommand{\arraystretch}{1.25}
\begin{tabular}{c|cc}
\hline
Variation     & $\sigma_{WW}^{\rm NLO}$ / $\sigma_Z^{\rm NLO}$  & $\sigma_{WW}^{\rm NLO}$ / $\sigma_Z^{\rm NNLO}$ \\
              & $\times 10^{-3}$     & $\times 10^{-3}$         \\
\hline
$r_s (-0.3)$	&	0.020	&	0.020	\\
$r_s (+0.3)$	&	-0.018	&	-0.019	\\
$M_c (-0.06$~GeV$)$	&	0.001	&	-0.001	\\
$M_c (+0.06$~GeV$)$	&	-0.002	&	-0.001	\\
$M_b (-0.45$~GeV$)$	&	0.000	&	0.002	\\
$M_b (+0.25$~GeV$)$	&	0.000	&	-0.001	\\
$\alpha_S(M_Z) (-0.002)$	&	0.011	&	0.010	\\
$\alpha_S(M_Z) (+0.002)$	&	-0.011	&	-0.010	\\
$Q^2_{\rm min} (-2.5$~GeV$^2)$	&	0.002	&	-0.007	\\
$Q^2_{\rm min} (+2.5$~GeV$^2)$	&	0.000	&	0.028	\\
$Q^2_0 (-0.2$~GeV$^2)$	&	0.000	&	-0.001	\\
+$D_{u_v}$	&	0.005	&	-0.015	\\
+$D_{\bar{U}}$	&	0.000	&	-0.001	\\
+$E_{\bar{U}}$	&	0.000	&	0.000	\\
\hline
\end{tabular}
\caption{\label{tab:tabratiomodel}Shifts of the ratios $\sigma^{\rm NLO}_{WW}/\sigma^{\rm NLO}_{Z}$ and  $\sigma^{\rm NLO}_{WW}/\sigma^{\rm NNLO}_{Z}$ 
due to the model and parameterisation variations in the PDF fit.
 }
\end{center}
\end{table}

An alternative approach to benefit from the partial cancellation of the 
PDF uncertainties is to use NNLO PDFs for the processes with only NLO matrix
element calculations. The mismatch of the calculation order is beyond the NLO
accuracy and thus could be considered to be covered by the NLO calculation 
uncertainty, which is estimated by the scale variation. 
Given the observed anti-correlations between NLO and NNLO sets,
this procedure may, however, lead to an underestimation of the PDF 
uncertainties.
A calculation of the $WW$ di-boson 
to $Z$ boson production cross-section ratio using the HF14cor-nlo-nnlo NNLO PDF set
yields
$$ 
\begin{array}{l}
\sigma_{WW}^{\rm NLO(NNLO\,PDF)}/\sigma_Z^{\rm NNLO} = \\[0.1in] =
[1.527\pm 0.008\,({\rm exp.})^{+0.023}_{-0.022}\,({\rm mod.})] \times 10^{-3}\,,
\end{array}
$$
where the uncertainties represent the experimental (exp.) and model plus 
parameterisation (mod.) PDF 
errors only, and are very similar to the PDF errors for the $\sigma_{WW}^{\rm NLO}/\sigma_{Z}^{\rm NLO}$ ratio.
The central value is consistent with  
the $\sigma^{\rm NLO}_{WW}/\sigma^{\rm NNLO}_{Z}$ calculation within $0.7\%$; 
however, the PDF uncertainties may be underestimated by $30-50\%$.

Adding the PDF and scale uncertainties (Table~\ref{tab:tabratiomodel})
in quadrature, the cross-section ratio of  $WW$ di-boson to $Z$ boson production 
calculated as the ratio of NLO predictions is 
$$ \sigma_{WW}^{\rm NLO}/\sigma_Z^{\rm NLO} = [1.543^{+0.073}_{-0.062}] \times 10^{-3}$$ 
and as the ratio of NLO to NNLO predictions is 
$$ \sigma_{WW}^{\rm NLO}/\sigma_Z^{\rm NNLO} = [1.517^{+0.051}_{-0.047}] \times 10^{-3} .$$
The usage of the mixed-order calculations
leads to a $30-40\%$ reduction of the overall theoretical uncertainty.

\section{Summary}
Sets of LO, NLO and NNLO parton distribution functions are reported preserving the correlations
of PDFs determined at different orders. The sets are determined with the HERAFitter program  using
the combined HERA data. The input parameters of the fits use recent experimental results on the 
charm-quark mass parameter $M_c$ and the strangeness suppression parameter 
$r_s$.
The experimental PDF uncertainties are determined using the MC method and
reported using both MC and eigenvector representations. A high degree of  correlation is 
observed for the PDFs at different perturbative order and similar Bjorken $x$. The model and parameterisation 
PDF uncertainties are estimated by varying the values of the input parameters and by adding extra
terms in the PDF parameterisation. 

The correlated NLO and NNLO PDF sets are used to calculate the $WW$ di-boson and $Z$ boson
production cross sections. The $WW$ di-boson production cross section is calculated at NLO
using MCFM. The $Z$ boson production cross section is calculated at  NLO and NNLO
using  FEWZ. Significant correlations of the PDF
uncertainties are observed for the cross sections calculated at different orders. For the ratio
of the $WW$ di-boson to $Z$ boson production cross sections an overall $30-40\%$ reduction of uncertainties is observed
when using mixed-order calculations due to the reduced higher order uncertainty for the $Z$ boson production 
cross section calculated at NNLO.

\begin{acknowledgements}
~\\
We are grateful to the DESY IT department for their support of the 
HERAFitter developers. 
This work is supported in part by  Helmholtz Gemeinschaft under
contract VH-HA-101, 
 BMBF-JINR cooperation program, ``Dinastiya'' foundation
 and RFBR grant
12-02-91526-CERN\_a.
\end{acknowledgements}
\bibliographystyle{utphys_mod} 
\bibliography{main}

\end{document}